\documentclass[prx, twocolumn, superscriptaddress, floatfix]{revtex4-2}
\usepackage{graphicx}
\usepackage{float}
\usepackage{amsmath,amssymb,amstext,dsfont,tikz,graphicx,physics,mathtools,bm,
simpler-wick}
\definecolor{mydarkred}{RGB}{165,4,66}
\usepackage{hyperref}
\hypersetup{
    colorlinks=true, 
    linkcolor=blue,  
    citecolor=mydarkred, 
    urlcolor=blue    
}

\newcommand{\ident}{\mathds{1}}

\newcommand{\SWAP}{\text{SWAP}}
\newtheorem{definition}{Definition}[section]

\begin{document}

\title{Generalizing measurement-induced phase transitions to information exchange symmetry breaking}
\author{Shane P. Kelly}
\email{skelly@physics.ucla.edu}
\affiliation{Mani L. Bhaumik Institute for Theoretical Physics, Department of Physics and Astronomy, University of California at Los Angeles, Los Angeles, CA 90095}
\affiliation{Institute for Physics, Johannes Gutenberg University of Mainz, D-55099 Mainz, Germany}
\author{Jamir Marino}
\affiliation{Institute for Physics, Johannes Gutenberg University of Mainz, D-55099 Mainz, Germany}

\date{\today}

\begin{abstract}
    Probing a quantum system disrupts its state in a phenomenon known as back action.
    In this work we investigate the conditions for this disruption to result in a phase transition in the information dynamics of a monitored system.
    We introduce a framework that captures a wide range of experiments encompassing probes comprised of projective measurements and probes which more generally transfer quantum information from the system to a quantum computer.
    The latter case is relevant to the recently introduced quantum-enhanced experiments in which quantum post-processing is performed on a quantum computer, and which can offer exponential sampling advantage over projective measurements and classical post-processing.
    Our framework explicitly considers the effects of an environment using a model of unitary evolution which couples system, apparatus and environment.
    Information dynamics is investigated using the R\'enyi and von Neumann entropies of the evolving state, and we construct a replica theory for studying these quantities.
    We identify the possible replica symmetries an experiment can possess and discuss the meaning of their spontaneous symmetry breaking.
    In particular, we identify a minimum subgroup whose spontaneous symmetry breaking results in an entanglement transition.
    This symmetry is only possible when the information in the apparatus is as informative about the dynamics of the system as the information transferred to the environment.
    We call this requirement the information exchange symmetry and quantify it by a relationship between the entropies.
    We then introduce a generalized notion of spontaneous symmetry breaking such that the entanglement transition can be understood as the spontaneous breaking of the information exchange symmetry and without referring to the replica theory.
    Information exchange symmetry breaking is then shown to generalize the phenomenology of the measurement-induced phase transition~(MIPT).
    We apply this theory to the brickwork quantum-enhanced experiment introduced in an accompanying letter~\cite{short} in the case where the unitaries are chosen from the Haar measure, and identify a distinct universality from the MIPT.
    This notion of information exchange symmetry breaking generalizes the MIPT and provides a framework for understanding the dynamics of quantum information in quantum-enhanced experiments.

\end{abstract}

\maketitle

\section{Introduction}
Back action is the phenomenon in which a measurement  of a quantum system changes its state and its subsequent dynamics.
Usually this effect is constrained to the microscopic domain, since at the macroscopic scale, the natural noise of the system dominates over effects due to quantum back action.
In contrast, the measurement induced phase transition~(MIPT) occurs in a setting that breaks this expectation by making projective measurements at a finite density and finite rate.
When such an extreme number of measurements are made, their microscopic effect has the possibility to collectively scale with the system size and the evolution time. 
Despite initial expectations that the measurements would simply destroy entanglement and quantum information~\cite{Chan_Nandkishore_OG}, two macroscopic phases of information dynamics can arise depending on the rate or density of measurements~\cite{Chan_Nandkishore_OG,LI_Fisher_OG,Skinner_Nahum_OG,fisherreview}.
Below a critical rate, the natural chaotic dynamics of a many body system spreads quantum correlations across the whole system, while above a critical rate those correlations become suppressed beyond a fixed length scale set by the measurement rate.
This was first demonstrated~\cite{Chan_Nandkishore_OG,LI_Fisher_OG,Skinner_Nahum_OG,fisherreview} in a model of chaotic quantum dynamics in which a one dimensional chain of qubits is evolved by the application of a random brickwork circuit.
After each step in the brickwork, each qubit is measured with a fixed probability $p$, such that an initial pure state evolves into a random pure state conditioned on the choice of random unitaries and measurement outcomes.

In contrast to a traditional phase transition,  random brickwork dynamics do not exhibit a unitary symmetry that could be spontaneously broken by the MIPT.
Thus, the phase transition is distinguished, not by a local order parameter, but by the structure of correlations as captured by various basis independent quantifiers such as the von Neumann entropy, R\'enyi entropy and Fisher information~\cite{Li_Fisher_2,Gullans_2020,theoryOftransitionsBao}.
The structure of correlations shows a phase transition both in equal time correlations such as the entanglement and purity, and in unequal time correlations such as in the quantum channel capacity between initial and final system state~\cite{Gullans_2020,Vija_Self_organizedCorrection,Altman_MIPT_Codes,scalable_probes,Li_fisher_stat_codes}.

This nature of the MIPT presents a serious problem for observing it: in the correlated phase, correlations are long range and non local such that detecting them requires the processing of an extensively growing number of measurements.
Initially~\cite{LI_Fisher_OG,Gullans_2020,Vija_Self_organizedCorrection,Altman_MIPT_Codes,scalable_probes,Li_fisher_stat_codes}, this was identified as a postselection problem in which late time observables of the phase transition  require conditioning on the outcomes of the measurements made during the dynamics.
Since each measurement has a random outcome, and there is an extensive number of measurements in both space and time, the probability of obtaining a repeated measurement is exponentially suppressed in the space-time volume of the experiment.
Significant effort has been put towards addressing this problem~\cite{noel_observation_2022,KohSuperConducting,Google_teleportation_MIPT_2023,ippoliti_postselection-free_2021,crossentropyOG,crossEntropyZ2sym,garratt_shadows,ScrambelingTransition}, ranging from experiments applying brute force approaches on small systems~\cite{noel_observation_2022,KohSuperConducting,Google_teleportation_MIPT_2023} to more elaborate proposals~\cite{crossentropyOG,crossEntropyZ2sym,garratt_shadows} which can address systems that can be simulated classically, or tools to probe dynamics in the uncorrelated phase~\cite{Google_teleportation_MIPT_2023,ippoliti_postselection-free_2021}.
Thus, while probes exist to study the uncorrelated~(area law) phase, it still remains to identify practical protocols that would confirm the predicted behavior in the correlated~(volume law phase) phase.

This work is motivated from two related directions aimed at understanding  the observable properties of the MIPT.
The first investigates the apparent requirement to post select the outcomes of projective measurements.
While the second aims at understanding the structure of the correlations and searches for efficient probes.
Previous attempts at understanding the requirement of postselection involve two approaches.
The first is studying adaptive dynamics in which the unitary dynamics of the system is driven with feedback from the past measurements~\cite{buchhold_preselection,Pixley_seperate_Feedback_2023_PRL,Steering-induced_2024,Vedika_comparing_MIPT_n_absorbing,hauser2023continuousbreaking,MIPTofMATTER}.
The second related approach is in identifying space-time correlations which might mimic the effect of a specific measurement sequence~\cite{finite_time_teleportation,ippoliti_postselection-free_2021,Grover_MBL_dualto_MIPT,ScrambelingTransition}.
Our approach encompasses both and investigates the requirements on the space-time structure of both the environment and measurement apparatus.

In order to do so, we construct a general framework which encompasses any type of environment and measurement apparatus.
In particular, we allow the measurement apparatus to perform not only projective measurements, but more generally quantum transduction.
Quantum transduction~\cite{Lauk_2020,LiangJian_SqueezingEnhnaced,LiangJiangOptimal} is a measurement process in which information about a system is obtained by directly transferring quantum information to a quantum storage device.
This is in contrast to projective measurements, in which the measurement apparatus transfers classical information to a classical storage device.
The importance of quantum transduction is that it allows for quantum post processing on the quantum information recorded by the measurement apparatus, and it has recently shown exponential advantage over experiments which perform classical postprocessing~\cite{quantumadvantage,boundsonlearning,aharonov2022quantum}.
The simplest example of quantum transduction is the swap gate which swaps a qubit of the interrogated system with a qubit in the quantum storage device.
Recently, such a swap operation was shown to induce an information phase transition, again on a random brickwork circuit, but where at a probability $p$ qubits are swapped to the quantum storage device~\cite{ScrambelingTransition}.
While that transition offers a decoder that can probe the phase transition throughout the phase diagram, it does not have the same phenomenology as the MIPT and in particular only shows up in the temporal correlations. 

Thus, in this work, and the accompanying letter~\cite{short}, we investigate the requirements on the system, environment and apparatus to show a transition in the structure of both temporal and spatial correlations similar to the MIPT.
We start by constructing a replica approach which has been useful in studying the critical properties of the MIPT~\cite{theoryOftransitionsBao,Nahum_all_to_all_Q_trees,Jian_Ludwig_criticality}.
Using this approach, previous work~\cite{theoryOftransitionsBao,Nahum_all_to_all_Q_trees,Jian_Ludwig_criticality} has equated the MIPT to a replica-symmetry-breaking transition.
Since replica symmetry breaking occurs as a symmetry breaking of a unitary symmetry, albeit on the unphysical replicas, it might have a local order parameter and thus offers a promising angle to identify practical probes of the transition.

\subsection{Overview of results}

After introducing key concepts and definitions in Sec.~\ref{SM:prelim}, and setting up the framework encompassing both classical and quantum-enhanced experiments in Sec.~\ref{SM:QEE}, we construct the replica theory in Sec.~\ref{SM:replica}.
Following approaches similar to those in Refs.~\cite{theoryOftransitionsBao,Nahum_all_to_all_Q_trees,Jian_Ludwig_criticality}, the replica theory introduces $n$ replicas of the system used to compute the $n^{th}$ R\'enyi entropy, and associated purities.
Then in Sec.~\ref{sec:rep_syms} we identify various forms of replica symmetry an experiment might possess and review the phenomena associated with the symmetry breaking of the different replica symmetries. 
These include the Anderson localization transition, in which local disorder tunes a transition in the conduction of non-interacting electrons, and the spin-glass transition, in which spins form random, but fixed, spin configurations below a critical temperature. 
In particular, we focus on the simplest case of $n=2$, relevant for the dynamics of purity, and we identify a necessary and sufficient condition for an experiment to possess the replica symmetry of the MIPT.

In Sec.~\ref{SM:SNEsym}, we investigate the physical meaning of this condition and identify ways to guarantee and probe it.
Specifically, we show it is related to an exchange of information between the apparatus and the environment.
We first identify the Local Unitary Exchange~(LUE) symmetry, acting on the system, apparatus and environment, and which guarantees the MIPT replica symmetry for $n=2$.
For $n\geq 2$, the LUE symmetry guarantees a $D_n\rtimes Z_2$ replica symmetry, and we show how this $D_n\rtimes Z_2$ symmetry subsequently guarantees a symmetry in the R\'enyi entropies.
Concretely, this symmetry in the R\'enyi entropies implies that the information in the apparatus is as useful as the information in the environment when determining properties about the system.
While the MIPT was initially discussed without reference to an environment, we show in Sec.~\ref{sec:projmeas} how projective measurements are equivalent to a unitary acting across the system, apparatus and an environment.
In doing so, we find that this information exchange~(IE) symmetry is present in the MIPT.

Of particular benefit, is that this IE symmetry does not require a replica theory to define and offers a potentially useful avenue to understanding the MIPT and related phase transitions.
Thus, in Sec.~\ref{sec:symcomp}, we compare it with global unitary symmetries, and in Sec.~\ref{sec:tasksym} we argue for it as a symmetry between communication tasks.
Later, in Sec.~\ref{sec:IEsb}, we show how the MIPT, and the phase transition occurring in the quantum-enhanced experiment introduced in the accompanying letter,~\cite{short} can be understood as the spontaneous breaking of the IE symmetry.

First, in Sec.~\ref{SM:numerics}, we reintroduce the random brickwork quantum-enhanced experiment studied in the accompanying letter~\cite{short}, and analyze it in the case of brickwork unitaries chosen from the full Haar measure.
Readers are suggested to read that letter before reading this section as it directly discusses the phenomenology of the phase transition and shows numerical results for unitaries chosen from the Clifford group.
In contrast, Sec.~\ref{SM:numerics} contains details on the Haar average important for understanding the replica symmetry breaking occurring in the MIPT.
Similar to Refs.~\cite{theoryOftransitionsBao,Nahum_all_to_all_Q_trees,Jian_Ludwig_criticality}, the average over the Haar measure induces a map to a statistical mechanics model of random graphs called the random bond cluster~(RBC) model.
For the finite $n$ replica theory, the RBC model, is equivalent to a Potts model of $h(n)$ spins, where $h(n)$ is a combinatorial factor determined by the $D_n\rtimes Z_2$ replica symmetry.
While the MIPT maps to directed percolation in the large local Hilbert space limit, we use this mapping to show that the random brickwork quantum-enhanced experiment falls within a different universality class.
This is consistent with the larger replica symmetry possessed by the $n>2$ replica theories for the MIPT, and the Clifford numerics discussed in the accompanying letter~\cite{short}.

While the MIPT and the random brickwork quantum-enhanced experiment fall within different universality classes, they show similar phenomenology to each other.
This is discussed in Sec.~\ref{sec:IEsb}, where we show that the spontaneous symmetry breaking of the $D_n\rtimes Z_2$ replica symmetry guarantees the same phase transition in the structure of correlations as captured by the R\'enyi Entropies.
This discussion is generalized beyond the replica theory by using the IE symmetry.
By developing the idea of IE symmetry breaking, we discuss how the von Neumann entropy can be used to observe the phase transition in the structure of correlations. 
This allows us to study the quantum conditional entropy, which unlike the R\'enyi entropies, satisfies the data processing inequality and allows for a direct discussion of quantum-entanglement for mixed states.
Doing so, we show that IE symmetry breaking is an entanglement transition~(Sec.~\ref{sec:enttrans}), a purification transition~(Sec.~\ref{sec:temptrans}) and a transition in quantum channel capacity~(Sec.~\ref{sec:temptrans} and Sec.~\ref{sec:tasksym}).
This completes our analysis of the IE symmetry breaking transition and in the discussion we summarize how it encompasses the MIPT, and its relation to a few other transitions discussed in the literature.
Finally, we conclude with prospects for future directions. 

\tableofcontents

\section{Preliminaries}~\label{SM:prelim}

In this section we introduce a few preliminaries required for describing a generic quantum experiment.
Our framework will capture all correlations relevant to the experiment and therefore needs the techniques of mixed state purification and channel to state isomorphisms.
These tools are reviewed in Sec.~\ref{sec:prelim_vec}, where we also introduce notation for their repeated use.
In the replica theory, permutation of replicas comprise the various replica symmetries.
Accordingly, we review the permutation group in Sec~\ref{sec:permgroup} and summarize standard notation for the group elements.
Finally, in Sec.~\ref{sec:haarprelim}, we review techniques for averaging over the Haar measure and introduce the permutation unitaries which are useful for discussing the replica symmetries.

\subsection{Vectorization and channel-state isomorphism}\label{sec:prelim_vec}

We consider an operator, $O$, that acts on a Hilbert space $\mathcal{H}$ of dimension $d$.
It can be written as a vector in the Hilbert space $\mathcal{H}\otimes \mathcal{H}^{*}$ where $\mathcal{H}^{*}$ is the dual space to $\mathcal{H}$.
This is accomplished using the unnormalized canonical maximally entangled state~\cite{nielsen2010quantum,choiiso}
\begin{eqnarray}
    \ket{\widetilde{\Phi_{\mathcal{H},\mathcal{H}^{*}}}}=\sum_{i=1}^{d}\ket{i,i}
\end{eqnarray}
The operator is written as a vector via:
\begin{eqnarray}\label{eq:vectorize}
    \ket{O}=O\otimes \ident\ket{\widetilde{\Phi_{\mathcal{H},\mathcal{H}^{*}}}}
\end{eqnarray}
Operators acting on this state obey
\begin{eqnarray}
    A\otimes B\ket{O}=\ket{AOB^{\intercal}}
\end{eqnarray}
where $B^{\intercal}$ is the transpose of $B$.
While the overlap of two vectorized operators is
\begin{eqnarray*}
    \braket{A}{B}=\tr\left[ A^{\dagger} B \right].
\end{eqnarray*}

Similarly, a linear map, $O\rightarrow O'=\mathcal{L}(O)$ on operators acting on states in $\mathcal{H}$ can be mapped to an operator acting on states in $\mathcal{H}\otimes \mathcal{H}^{*}$.
This is generally known as the Choi–Jamiołkowski isomorphism~\cite{choiiso}. 
Specifically, we write the Choi matrix as:
\begin{eqnarray}
    \sum_{ij}\mathcal{L}(\ket{i}\bra{j})\otimes\ket{i}\bra{j}.
\end{eqnarray}
When the map is a CPTP quantum channel, then the Choi matrix is positive.
When the map is unitary conjugation, $\mathcal{L}(O)=VOV^{\dagger}$, that Choi matrix has the form of the pure state density matrix
\begin{eqnarray}
    \ket{V}\bra{V}
\end{eqnarray}
where $\ket{V}$ is the vectorization, Eq.~\eqref{eq:vectorize}, of the unitary $V$.

\subsubsection{State purification}~\label{sec:statepurification}

A general mixed state $\rho$ acting in a Hilbert space $\mathcal{H}$ can always be written as a pure state in a larger space $\mathcal{H}_1\otimes \mathcal{H}_2$, via the vectorization of the operator $\rho^{1/2}$:
\begin{eqnarray}
    \ket{\sqrt{\rho}}=\sqrt{\rho}\otimes 1 \ket{\widetilde{\phi_{\mathcal{H}_1,\mathcal{H}_2}}},
\end{eqnarray}
such that $\rho=\tr_2\left[ \ket{\sqrt{\rho}}\bra{\sqrt{\rho}} \right]$.

\subsection{Average of unitaries chosen from the Haar measure}~\label{sec:haarprelim}
The Haar average over the matrix elements of $n$ unitaries and their conjugates is given as:
\begin{eqnarray}\label{eq:wgaverage} 
    &\overline{\prod_{k=1}^{n}U_{i_k j_k}\prod_{l=1}^{n}U^{*}_{m_l r_l}}=\\ \nonumber
    &\sum_{\tau,\sigma\in S_n}W_{g}^{d,n}(\tau,\sigma)\prod_{k=1}^{n}\delta(i_k=m_{\sigma(k)})\delta(j_k=r_{\tau(k)})
\end{eqnarray}
where the sum over $\tau$ and $\sigma$ is a sum over all permutations in the permutation group $S_n$, and $W_g^{d,n}$ is the Weingarten function~\cite{Weingarten}.
This can be written efficiently, by defining the permutation unitaries
\begin{eqnarray}
    X_p^{(n)}=\sum_{\{x_i\}}\ket{x_{p(1)},x_{p(2)},\dots x_{p(n)}}\bra{x_1,x_2,\dots x_n},
\end{eqnarray}
and their vectorization $\ket{\sigma} = X_{\sigma}^{(n)}\ket{\widetilde{\Phi_{H,H^*}}}$.
This expression is then written as:
\begin{eqnarray}~\label{eq:WGdef}
    \overline{U^{\otimes n}\otimes U^{*\otimes n}}=W_{g}^{d,n}=\sum_{\tau,\sigma\in S_n}W_{g}^{d,n}(\tau,\sigma)\ket{\sigma}\bra{\tau}
\end{eqnarray}
Below, we will refer to specific elements of the permutation group for $\tau$ and $\sigma$.
See the appendix~\ref{sec:permgroup} for a review of the standard notation used for permutation groups.

\subsubsection{Permutation unitaries}
The vectorized permutation unitary satisfies the following useful properties.
Under application of a permutation unitary, it transforms as
\begin{eqnarray}
    X^{(n)}_x\otimes X^{(n)}_{y}\ket{\sigma}=\ket{x\sigma y^{-1}}.
\end{eqnarray}
Second is its overlap with pure states $\ket{\psi^{n}}=\ket{\psi}^{\otimes n}\otimes \ket{\psi}^{*\otimes n}$.
\begin{eqnarray}
    \braket{\psi^{n}}{\tau}=\braket{\psi}{\psi}^{n}
\end{eqnarray}
Finally, the overlap between two permutation unitaries,
\begin{eqnarray}
    Q^{d,n}(\tau,\sigma)=\braket{\tau}{\sigma}=d^{\text{\# cycles}(\tau\sigma^{-1})},
\end{eqnarray}
is the inverse of the Weingarten function
\begin{eqnarray}
    \delta(\sigma, \tau)=\sum_{p}Q^{d,n}(\sigma,p)W_g^{d,n}(p,\tau).
\end{eqnarray}

\section{Framework for quantum-enhanced experiments}~\label{SM:QEE}
In this section we extend the framework of an experiment presented in the accompanying letter~\cite{short} to allow for the possibility of long range temporal correlations in the environment and apparatus, and to accommodate arbitrary initial states.
To completely characterize how information evolves, we construct a picture involving the explicit purification of mixed states and environments.
In this way, we treat the dynamics of all relevant components of the experiment by a global unitary.
Then, to capture temporal correlations, we use the channel to state isomorphism  discussed above.
We first motivate and introduce the relevant Hilbert spaces, then describe the initial state purifications, and finally present the channel-state isomorphism.
This construction results in a pure quantum state which captures all correlations between the relevant points of space-time involved in the experiment.

\begin{figure}[t]
    \includegraphics[width=1\columnwidth]{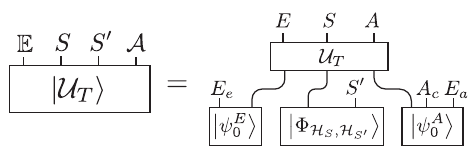}
    \caption{Model for a generic experiment, classical or quantum-enhanced, in which the world is divided into the system of interest $S$, the apparatus used to observe the system $\mathcal{A}=A\cup A_c$, and the environments $\mathbb{E}=E\cup E_a\cup E_e$ which are a source of noise. The $E$ and $A$ components of the environment and apparatus interact with the system $S$ by the unitary $\mathcal{U}_T$. The environment component $E$ and both the apparatus components $\mathcal{A}=A\cup A_c$ are initialized in the mixed states $\rho^E_0=\tr_{E_e}[\ket{\psi^{E}_0}\bra{\psi^{E}_0}]$ and  $\rho^A_0=\tr_{E_a}[\ket{\psi^{A}_0}\bra{\psi^{A}_0}]$. To treat the temporal and spatial correlations on the same level, we introduce an auxiliary system $S'$ and perform the channel to state isomorphism with the maximally entangled state $\ket{\Phi_{\mathcal{H}_S,\mathcal{H}_{S'}}}$. The resulting ``Choi'' state $\ket{\mathcal{U}_T}$ of the channel is shown.}
	\label{fig1}
\end{figure}

We consider a class of experiments that probes the dynamics of a system of interest.
The system is coupled to some environment, and the experiment proceeds by intermittently coupling a measurement apparatus to the system.
In the most general case~(shown in Fig.~\ref{fig1}), one may consider the environment, system and apparatus all undergoing some global unitary evolution, $\mathcal{U}_T$.
Noise can come from both the initial and final states of the environment and apparatus.
When the final state of the environment is ignored, information is lost and introduces noise into the system and apparatus.
Similarly, when the initial state of the environment or apparatus is mixed, noise is injected into the system and apparatus.
Furthermore, while it is obvious useful information is obtained from the final state of the apparatus, the initial state can also act as a valuable source of information.
An extreme case of this is in the Hayden-Preskill protocol~\cite{PatrickHayden_2007,yoshida2017efficient}, in which a diary thrown into a black hole can be recovered if the experimenter has full control of the black hole initial state, and only a small fraction of the Hawking radiation.  
Similar to this protocol, we imagine that the qudits of the apparatus can be prepared in a maximally entangled state with an auxiliary set of qudits.
This copy of the initial state of the apparatus is then an additional source of quantum information available to the experimentalist.

In order to treat these different sources of informaiton and noise on a single footing, we work in the framework of a pure state evolving by unitary evolution.
We assume that the physical space of the system does not change over the duration of the experiment such that the initial state of the system exists in a Hilbert space $\mathcal{H}_{S'}$ of the same size as the Hilbert space of the final system states, $\left|\mathcal{H}_{S}\right|=\left|\mathcal{H}_{S'}\right|$.
Furthermore, we assume that the system has a product space structure $\mathcal{H}_S=\otimes_{i\in S}\mathcal{H}_{S_i}$. 
Concretely, we discuss examples in which the local Hilbert spaces $\mathcal{H}_{S_i}$ correspond to qudits, but they could more generally be different modes or quanta of a radiation field, such as the photons emitted from a black hole under Hawking radiation.

We model the dynamics of the experiment as a unitary which evolves states in the system, apparatus and environment, $\mathcal{H}_{E'}\otimes\mathcal{H}_{A'}\otimes\mathcal{H}_{S'}$, to states in $\mathcal{H}_E\otimes\mathcal{H}_A\otimes\mathcal{H}_S$, where $E'$, and $A'$ label the initial environment and apparatus spaces, while $E$ and $A$ label the respective final spaces.
We also allow the environment and apparatus to exchange quanta, such that $|\mathcal{H}_A|\neq|\mathcal{H}_{A'}|$, but under the constraint that the global Hilbert space remains the same $|\mathcal{H}_A\otimes\mathcal{H}_E|=|\mathcal{H}_{A'}\otimes\mathcal{H}_{E'}|$.
We assume that the initial state of the environment is in general some mixture $\rho^{E}_0$ of states living in $\mathcal{H}_{E'}$, but introduce an auxiliary environment $\mathcal{H}_{E_e}$ to purify~(See section~\ref{sec:statepurification}) the mixture $\rho^{E}_0=\tr_{E_e}[\ket{\psi^{E}_0}\bra{\psi^{E}_0}]$.
In this way, we can explicitly keep track of the noise introduced by the uncertainty in the initial environment states.

Finally, we consider the experimenter to be able to store qudits in a Hilbert space $\mathcal{H}_{A_c}$ which doesn't directly couple to the system.
In this way the experimenter can prepare an initial state $\rho_{0}^{A}$ which may contain entanglement between $\mathcal{H}_{A'}$ and $\mathcal{H}_{A_c}$.
If such entangled states are the canonically maximally entangled state, the state of $\mathcal{H}_{A_c}$ act as a quantum record of the initial states directly coupled to the system.
Such entangled states are needed for channel transduction~\cite{Lauk_2020,quantumadvantage,choiiso}, and for the Hayden-Preskill protocol for black hole decoding~\cite{PatrickHayden_2007,yoshida2017efficient}.
Again, we directly consider the noise introduced by uncertainty in the initial apparatus state and introduce the purification $\ket{\psi^{A}_0} \in \mathcal{H}_{A'}\otimes\mathcal{H}_{A_c}\otimes \mathcal{H}_{E_a}$ of the state $\rho^A_0=\tr_{E_a}[\ket{\psi^{A}_0}\bra{\psi^{A}_0}]$

In this way we have divided the universe into a global environment $\mathbb{E}=E\cup E_a\cup E_e$ which captures all source of noise, the system $S$, and the information recorded on the quantum computer $\mathcal{A}=A \cup A_{c}$.
When disregarding the environments, the experiment implements a quantum channel from $\mathcal{H}_{S'}$ to $\mathcal{H}_\mathcal{A}\otimes\mathcal{H}_S$.
To treat the initial and final state of the system in an equal way we use the Choi–Jamiołkowski isomorphism~(See sec~\ref{sec:prelim_vec}) to map this channel to the state $\rho_\mathcal{U}$.
When including the environments, we can consider the purification of this Choi state
\begin{eqnarray}
    \ket{\mathcal{U}_T}=\mathcal{U}_T\ket{\psi^\mathcal{A}_0}\ket{\psi^\mathbb{E}_0}\ket{\Phi_{\mathcal{H}_S,\mathcal{H}_{S'}}}
\end{eqnarray}
such that $\rho_{\mathcal{U}}=Tr_{\mathbb{E}}[\ket{\mathcal{U}_T} \bra{\mathcal{U}_T}]$.
This state represents a complete model of the dynamics of an experiment for any initial system state, whether it is mixed or pure. 
For notational convince, we introduce the final state of the system and apparatus when the system is initialized in a fixed pure state $\ket{\psi}$.
\begin{eqnarray}
    \ket{\mathcal{U}_T(\psi)}=\mathcal{U}_T\ket{\psi^\mathcal{A}_0}\ket{\psi^\mathbb{E}_0}\ket{\psi}.
\end{eqnarray}

\subsection{Markovian Environment and Apparatus}
The above framework of a quantum-enhanced experiment captures a large class of dynamics including, both Markovian and non-Markovian open quantum systems, hybrid circuits composed of projective measurements and unitaries~\cite{LI_Fisher_OG,Chan_Nandkishore_OG,Skinner_Nahum_OG,Li_Fisher_2,fisherreview,Gullans_2020,Vija_Self_organizedCorrection,Altman_MIPT_Codes,scalable_probes,Li_fisher_stat_codes,CritMIPT_Pixley,Sierant_Turkkeshi_Multi_Fracality,Nahum2023RenormalizationGF,zabalo_operator_multi_fractality,infinite_randomness,QuantumTreesNahum,theoryOftransitionsBao,Nahum_all_to_all_Q_trees,Jian_Ludwig_criticality,Li_Fisher_open_system,Zack_powerlaw_negativity,Piroli_Sym_resolved_page_curve,noel_observation_2022,Google_teleportation_MIPT_2023,KohSuperConducting,finite_time_teleportation,ippoliti_postselection-free_2021,Grover_MBL_dualto_MIPT,crossentropyOG,crossEntropyZ2sym,garratt_shadows,ippoliti_learnability_shadows,agrawal2023observing,mcginley2023postselectionfree,fux2023entanglementmagic,bejan2023dynamical,niroula2023phase,ICTP_Fazio_Diagramatics_nonMarkov,ICTP_Giuliano_nonMarkov_Free_Fermion,Vija_involumelaw,MeasurementProtected_Hsieh_Timothy,Bao_sym_enriched,Lavasani_2021_sym,Lavasani_2p1,Sarang_ChargeSharp2022PRL,charge_sharpening_Halpern_2023,barratt_sharpening_Learnability_PRL,agrawal_charge_sharpening_PRX,Poboiko_Mirlin_replica_theory,poboiko2023measurementinduced,pöpperl2024localization,Buchhold_free_fermions_PRL,Buchold_free_fermions_PRX,Buchhold_long_ranged,Giulia_Russomanno_issing,Turkeshi_Fazio_Free_fermion_zero_clicks,Turkeshi_Piroli_neagtivity_fermions,szyniszewski_lunt_arijeet_disordered_fermions,Turkeshi_2022_stocastic_reseting,LongRange_Ehud_Yao,Turkeshi_Fazio_Long_ranged,Russomanno_Giulia_2023,Turkeshi_Fazio_2p1,Lunt_relation_percolation_2p1,Sierant_Turkeshi_dp1,KPZ_classical_walker,Pizzi_brdiging_classical_QM_gap,Knolle_classical_chaotic_MIPT,Lyons_altman_classical_crossover,CoherenceBoundsKelly,Sierant_Fazio_Turkeshi_trapped_ion,Yuto_Yohei_continuous_monitoring,Mirlin_under_ancilla,martínvázquez2023phase,OG_monitored_fermions,Hafezi_continuous_monitoring,schiro_measured_subradiance,liu2024entanglement,liu2024noiseinduced}, unitary dynamics with measurements and feedback~\cite{Pixley_seperate_Feedback_2023_PRL,Steering-induced_2024,EntSteering,Vedika_comparing_MIPT_n_absorbing,buchhold_preselection,hauser2023continuousbreaking,MIPTofMATTER,Piroli_TrivialityQTcloseDP,Piotr_EntVAbs,sierant_EntVAbsdp1}, and the hybrid circuit of unitaries and swaps of the recently proposed scrambling transition~\cite{ScrambelingTransition,SYKscrambeling}.
While our discussion of the information exchange symmetry will hold for the general framework introduced above, the specific examples we consider are for experiments in which the system interacts with uncorrelated parts of the apparatus and environment at different times.
In these examples, the apparatus and environment cannot introduce temporal correlations to the system such that they are both considered Markovian with respect to the system~\cite{ModiProcessTensorOG}.
In the modern framework of non-Markovian quantum processes~\cite{ModiProcessTensorOG}, these experiments have a process tensor describing the system environment dynamics which factorizes, and a quantum comb~\cite{QuantumCombOG} describing the coupling of the system and apparatus which is uncorrelated and also factorizes.

\begin{figure}[t]
    \includegraphics[width=1\columnwidth]{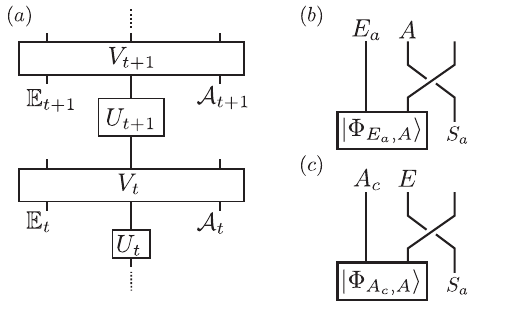}
    \caption{(a) Structure of the Markovian experiment. (b) Transduction with initially maximally mixed apparatus. (c) Information exchange dual to operation shown in (b). In both operations, the status of the apparatus component that directly couples to the system changes in time. In (b) the initial state is inaccessible to an experiment, but its final state is. In contrast, in (c) the experiment has access to the initial state but not the final state. Physically, the first case might correspond to an experiment in which part of the apparatus is initially at a higher temperature, while the second might correspond to loss of quantum information in part of the apparatus due to noise. In both cases coherent experimental control over the Hilbert space initially labeled $A$ is assumed at intermediate times.}
	\label{fig2}
\end{figure}

Figure~\ref{fig2} shows a generic model for such a Markovian experiment.
In that model, the environment, $\mathbb{E}$, and apparatus $\mathcal{A}$, are broken in to $T$ components $\mathbb{E}_t=E_t\cup E_{e,t}\cup E_{a,t}$ and $\mathcal{A}_t=A_t\cup A_{c,t}$, and involves $T$ steps.
Each step occurs with first the natural unitary dynamics, $U_t$, of the system, followed by another unitary $V_t$ coupling the environment $E_t$ and apparatus component $A_t$.
In addition, we require the initial states of the environment and apparatus, $\rho^{E}_0$ and $\rho_{0}^{A}$, to factorize across all $T$ components, such that initial conditions cannot introduce temporal correlations within the system.
We will also allow for distinct unitaries, $U_{t}$ and $V_t$, to occur at different times.
This setup is sufficiently general, that if a continuum limit is taken, and the unitaries $U_t$ and $V_t$ are chosen property, it can model Lindblad dynamics interspersed with coupling to an apparatus.

We now introduce four probe operations $V_t$ that have been discussed in the literature.
The first two are types of quantum transduction~\cite{Lauk_2020,LiangJian_SqueezingEnhnaced,LiangJiangOptimal} which have been shown to allow for an exponential advantage in quantum-enhanced experiments~\cite{quantumadvantage,boundsonlearning,aharonov2022quantum}.
The third is a unitary implementing projective measurements, while the forth is the noisy-transduction operations introduced in the accompanying letter~\cite{short} and studied in detail below.

\subsubsection{Transduction of a single qubit}
In Fig.~\ref{fig2} b), we show the operation of transducing a single qubit of the system into an initially maximally mixed apparatus.  
It simply swaps the qubit of the system with the qubit of the apparatus. 
In the figure, we have also depicted the purification of the maximally mixed state using the environment $E_a$.
This operation was used in Ref.~\cite{ScrambelingTransition} to compose a brickwork circuit interspersed with single qubit transduction occurring at a rate $p$.
Such a circuit shows a transition in the spreading of operators within the system and apparatus.
Above a critical rate $p_c$, an operator initially in the system is swapped into the apparatus after a fixed number of steps of the circuit.
While below the critical rate $p_c$, the operator maintains support on the system for times scaling exponentially in system size.
This transition, which we will refer to as the ``Scrambling Transition'', does not have the same phenomenology as the MIPT or the entanglement transition presented here.
This is because the operation shown in Fig.~\ref{fig2} b) does not satisfy the IE symmetry and so it cannot obey a MIPT replica symmetry~(See below in Sec~\ref{SM:SNEsym}).

The operation that results upon the exchange of the apparatus and environment is instead the one shown in Fig.~\ref{fig2} c).
In that case, the system qubit ejected from the system is erased, while the one injected is entangled with the apparatus.
The resulting circuit becomes very similar to the Hayden Preskill protocol for a black hole undergoing continuous hawking radiation and absorption of matter.
Using the purity of the global state, $\ket{\mathcal{U}}$, it can be shown that the scrambling transition is also a transition in the capacity to perform the Hayden Preskill protocol.
This is shown in a related work~\cite{gribben2024markovian} on a transition between Markovian and non-Markovian dynamics.

\subsubsection{Projective measurements}~\label{sec:projmeas}

In most experiments, the coupling of the apparatus implements a projective measurement and results in a measurement outcome recorded on a classical bit.
It has long been known that such a process can be captured by a unitary acting on an apparatus and environment~\cite{nielsen2010quantum}.
As discussed in the accompanying letter~\cite{short}, and shown in Fig.~\ref{fig:PM_NT} a), the projective measurement can be implemented by two CNOT gates, one coupling the measured bit to the apparatus and the other to the environment.
If the initial state of the system qubit was $\ket{\psi}=\psi_0\ket{0}+\psi_1\ket{1}$, the mixed state on the system and apparatus will be $\rho = \left|\psi_0\right|^2 \ket{00}\bra{00}+\left|\psi_1\right|^2 \ket{11}\bra{11}$.
Notice that this statistical distribution of quantum states involves the uncertainty of the measurement outcome, and is therefore not a pure state.

This is distinct from typical treatments of projective measurements, in which the projective measurement results in the system bit being in a pure quantum state. 
This pure state is the state in the system, conditioned on the outcome observed in the system.
For example, if the experimenter observes a `$0$' in the measurement apparatus, they know the system is in the pure state $\ket{0}=\ket{0}\braket{0}{\psi}/\braket{0}{\psi}$.
In both cases, the measurement outcomes and physical operations are the same.
The difference is the distribution of quantum states considered: In the former case, we consider the joint distribution on both system and apparatus, while in the latter, the conditional distribution is considered, conditioned on a specific measurement outcome.

This framework of measurements was used in Ref.~\cite{theoryOftransitionsBao} to construct a replica theory similar to our approach below. In Ref.~\cite{MIPTofMATTER}, the authors elaborate on this construction for studying the MIPT~(or MIET in their language) using the formalism of the Stinesping~\cite{Stinespring1955PositiveFO}. We discuss further connections to their work in the discussion.

\subsubsection{Noisy Transduction}~\label{sec:noisy_transduction}
\begin{figure}[t]
    \includegraphics[width=2.8362in]{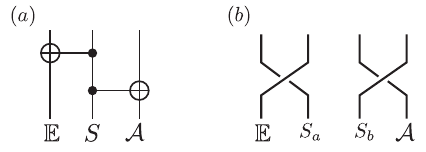}
\caption{Two examples of unitary couplings between the system, apparatus, and environment. (a) Projective measurements are implemented by two CNOT gates which only transfer information about the computational basis into the apparatus. (b) Noisy transduction transfers the full quantum state of qubit $S_b$ into the apparatus.}
	\label{fig:PM_NT}
\end{figure}

In this work, and in the accompanying letter~\cite{short}, we introduce the noisy-transduction operation.
This operation is shown in Fig.~\ref{fig:PM_NT} b) and involves the single qubit transduction of one qubit, and the quantum bit eraser of another qubit.
Both operations are modeled via a swap gate, one swapping the qubit into the apparatus, and the other swapping the qubit into the environment.
In the accompanying letter, we considered the initial state of the environment and apparatus to be an arbitrary pure state.
In this article, we will also briefly consider an experiment which conditions on the initial state of the apparatus qubit, and an environment qubit which is maximally mixed.
In both cases, the IE symmetry is satisfied so long as the system dynamics are symmetric under exchange of the two qubits.

\section{Replica Symmetry in Quantum Enhanced Experiments}~\label{SM:replica}

In this work we consider the dynamics of quantum information using three different sets of entropy quantifiers.
The first set is the $n^{th}$ purities defined as
\begin{eqnarray}
    \mu^{(n)}=\tr[\rho^n],
\end{eqnarray}
the second set are the $n^{th}$ R\'enyi entropies,
\begin{eqnarray}
    S^{(n)}(\rho) =\frac{1}{1-n}\log(\tr[\rho^n]),
\end{eqnarray}
while the last is the von Neumann entropy
\begin{eqnarray}
    S(\rho)=-\tr[\rho\log(\rho)]=\lim_{n\rightarrow 1}S^{(n)}(\rho).
\end{eqnarray}
The $n^{th}$ purities and R\'enyi entropies can be investigated using a replica approach, while the von Neumann entropies are obtained by the $n\rightarrow 1$ replica limit.

The replica approach represents these nonlinear quantities of the density matrix, such as $\tr[\rho^n]$, as linear observables of a replicated state $\rho^{\otimes n}$.
After disorder averaging or conditioning on the state of the apparatus, a tensor on the $n$ replica Hilbert spaces can be used to compute various entropy quantifiers.
Such a tensor can possess various replica symmetries depending on the dynamics of interest.
In this work we introduce a new replica symmetry that generalizes the one spontaneously broken in the MIPT~\cite{Buchold_free_fermions_PRX,theoryOftransitionsBao,Nahum_all_to_all_Q_trees,Jian_Ludwig_criticality}.

First, in Sec.~\ref{sec:replicareview} we review the replica symmetry and its spontaneous symmetry breaking, but in the more general context of the class of experiments introduced above.
In this case, we find that the replica symmetry, spontaneously broken in the MIPT, is generally not present for a generic quantum-enhanced experiment.
Instead, we find that a subgroup of the replica symmetry can persist in a generic experiment, but only when certain conditions are satisfied.
In Sec.~\ref{sec:n2sym}, we identify, within the replica theory, what these conditions are and the symmetry of the resulting subgroup.
Then, in the following Sec.~\ref{SM:SNEsym}, we show that the physical meaning of this condition is as a symmetry between the information in the environment and the information in the apparatus.

As we will show, both the replica symmetry and the information exchange symmetry are symmetries of the correlations in the Choi state $\ket{\mathcal{U}_T}$, describing the unitary evolving the system, apparatus and environment.
To quantify these correlations, we study the R\'enyi entropies 
\begin{eqnarray}~\label{eq:PSnSpdef}
    S^{(n)}(P_S,P_{S'};M)=S^{(n)}(\tr_{P_S^c\cup P_{S'}^{c}\cup M^c}[\ket{\mathcal{U}_T}\bra{\mathcal{U}_T}])
\end{eqnarray}
which give the $n^{th}$ R\'enyi entropy for the reduced density matrix on the subsystems $P_S\cup P_{S'}\cup M$, where $P_S\subset S$ is a subset of the final system qudits, $P_{S'}\subset S'$ is a subset of the initial system qudits, and $M\subset \mathcal{A} \cup \mathbb{E}$ is a subsystem of the apparatus and environment.
Here, we also define the complement $P_S^c$ of the subset $P_S$ as to be taken within the set $S$ such that $P_S \cup P_S^c = S$.
We define the complement similarly for $P_{S'}$ and $M$, where their complements are within there respective supersets~(i.e. $P_{S'}^c\cup P_{S'}=S'$). 
Similarly, we also investigate properties of the von Neumann entropies $S(P_S,P_{S'};M)=\lim_{n\rightarrow 1}S^{(n)}(P_S,P_{S'};M)$.

The notation $S(P_S,P_{S'};M)$ is introduced to reflect the structure of the experiments under consideration. The first argument will only contain subsets of the final system qubits, the second will only contain subsets of the initial system qubits, and the final argument will be either the environment or the apparatus qudits.  Since we often consider entropies conditioned on the final argument, we separate this argument by a semicolon.
Specifically, we consider the conditional R\'enyi entropies 
\begin{eqnarray*}
    S^{(n)}(P_S,P_{S'}|M) = S^{(n)}(P_S,P_{S'};M) - S^{(n)}(\varnothing,\varnothing;M)
\end{eqnarray*}
which describe the dynamics of information conditioned on the quantum state recorded in $M$.
Finally, we refer to the entropies for a fixed initial pure state, $\ket{\psi}$ as:
\begin{eqnarray*}
    S^{(n)}_{\psi}(P_S;M)=S^{(n)}(\tr_{P_S^c\cup M^{c}}[\ket{\mathcal{U}_T(\psi)}\bra{\mathcal{U}_T(\psi)}])
\end{eqnarray*}
and similarly for the conditional entropies $S^{(n)}_{\psi}(P_S|\mathcal{A})$.

To discuss the replica symmetry, we take the following approach.
First, we introduce a \textit{conditional replica tensor}, $\Sigma^{(n)}$, which acts on the replicated system space, $\left(\mathcal{H}_S\otimes\mathcal{H}_{S'}\right)^{\otimes n}$, and allows evaluating $S^{(n)}(P_S,P_{S'};\mathcal{A})$ as linear observables on $\Sigma^{(n)}$ for arbitrary $P_S$ and $P_{S'}$ and fixed $\mathcal{A}$.
Next we derive a Markovian kernel $\mathcal{K}^{(n)}$ which determines how $\Sigma^{(n)}(t)$ evolves after each step in the Markovian quantum experiment shown in Fig.~\ref{fig2}.
In the case of no apparatus, the kernel is a completely positive trace preserving~(CPTP) map acting on the replicated space $(\mathcal{H}_S\otimes H_{S'})^{\otimes n}\otimes(\mathcal{H}_S^{*}\otimes H_{S'}^{*})^{\otimes n}$.
We now review the replica approach, introduce these tensors, and discuss the replica symmetry possessed by various experiments.

\subsection{Review of the replica approach}~\label{sec:replicareview}

The replica approach represents nonlinear functions of matrices acting on one Hilbert space $\mathcal{H}$ of dimension $d$ such as $\tr[\rho^n]$ or $\tr[B^jA^kC^m]$, as linear observables on a replicated Hilbert space $\mathcal{H}^{\otimes n}=\mathcal{H}^{\otimes (j+k+m)}$.
The linear observables which accomplish this are all permutation unitaries 
\begin{eqnarray}
    X_p^{(n)}=\sum_{\{x_i\}}\ket{x_{p(1)},x_{p(2)},\dots x_{p(n)}}\bra{x_1,x_2,\dots x_n}
\end{eqnarray}
where the sum is over the basis states of $\mathcal{H}^{\otimes n}$~(i.e. $x_{i}\in [1,d]$), and the function $p(i) \in [1,n]$ is a one-to-one function of the indices $i\in[1,n]$.
The set of all functions $p(i)$ is the permutation group $S_n$, and the operators $X_p^{(n)}$ are a representation of the permutation group acting on $\mathcal{H}^{\otimes n}$.

Expectations of a replicated density matrix, $\rho^{\otimes n}$ for a given permutation unitary $X_p^{(n)}$ yield 
\begin{eqnarray}
    tr[X_p^{(n)} \rho^{\otimes n}] = \prod_{c\in \text{cycles}(p)}\tr[\rho^{\left|c\right|}]
\end{eqnarray}
where $\text{cycles}(p)$ gives the cycles in the permutation, and $|c|$ gives the length of the cycle.
Thus any $n$-cycle $p$ gives $tr[X_p^{(n)} \rho^{\otimes n}]=\tr[\rho^{n}]$, and for most uses, we will use the shift permutation $X_{+1}^{(n)}$ as the $n$-cycle of choice.
Similarly for the identity permutation $X_\mathds{1}=1$, we find $\tr[X_{\mathds{1}}\rho^{\otimes n}]=\tr[\rho^{\otimes n}]=\tr[\rho]^{n}$.

While we will not need it, it is interesting to note that more generally
\begin{eqnarray}
    tr[X_p^{(n)} \otimes_i^{n} A_i] = \prod_{c\in \text{cycles}(p)}\tr[T_c\prod_{j\in O(c)}A_j]
\end{eqnarray}
where $T_c$ enforces cycle order, and $O(c)$ gives the orbit of the cycle (i.e. the elements visited in the given cycle).

As a simple and concrete example, take $n=2$, for which the permutation group of two elements is $Z_2$ ~($\SWAP=X^{(2)}_{+1}$ and $\SWAP^2=\mathds{1}$).
In this case, the purity of a normalized density matrix can be written as:
\begin{eqnarray}
    \tr[\rho^2]=\tr[X_{+1}^{(2)} \rho \otimes \rho] 
\end{eqnarray}
while the normal of an operator is simply $\tr[A]=\sqrt{\tr[\mathds{1} A \otimes A]}$.

\subsection{The conditional replica tensor}

Using the above formalism, we construct the \textit{unnormalized conditional replica tensor} 
\begin{eqnarray}
    \widetilde{\Sigma}^{(n)} = \tr_{\mathcal{A} \cup \mathbb{E}}[(\ket{\mathcal{U}_T}\bra{\mathcal{U}_T})^{\otimes n}X_{+1}^{\mathcal{A}}]
\end{eqnarray}
where we define $X_{+1}^{\mathcal{A}}$ as the cyclic shift permutation unitary acting on the Hilbert space $\mathcal{H}_{\mathcal{A}}^{\otimes n}$.
The entropies of interest can then be obtained by linear expectation values of the permutation unitaries:
\begin{eqnarray}
    S^{(n)}(P_S,P_{S'};\mathcal{A}) = \frac{1}{1-n}\log(\tr[ X_{+1}^{P_S\cup P_{S'}}\widetilde{\Sigma}^{(n)}])
\end{eqnarray}
Furthermore, the conditional R\'enyi entropies are obtain in a similar fashion on the normalized conditional replica tensor, $\Sigma^{(n)}=\widetilde{\Sigma}^{(n)}/\tr[\widetilde{\Sigma}^{(n)}]$,
\begin{eqnarray}
    S^{(n)}(P_S,P_{S'}|\mathcal{A}) = \frac{1}{1-n}\log\left(\tr\left[ X_{+1}^{P_S\cup P_{S'}}\Sigma^{(n)}\right]\right)
\end{eqnarray}

Finally, we describe how to obtain the conditional entropies for the system initialized in either a fixed pure state, $\ket{\psi}$, or a maximally mixed state, $\rho_{\infty}=\mathds{1}/d$.
For a maximally mixed initial state, $\rho_{\infty}$, the entropies of interest, $S^{(n)}(P_S,\varnothing|\mathcal{A})$, are obtained from the reduced conditional replica tensor on the final system $S$~(i.e. $\Sigma^{(n)}_{\infty}=Tr_{S'}[\Sigma^{(n)}]$).
This is because $\Sigma^{(n)}$ was created by applying $\mathcal{U}_T$ to a maximally entangled state between $S$ and $S'$, such that discarding $S'$ is equivalent to injecting a maximally mixed state.
Instead, considering the entropies for a initially pure state $\ket{\psi}$, the reduced tensor $\widetilde{\Sigma}^{(n)}_\psi=Tr_{S'}[(\ket{\psi}\bra{\psi}_{S'})^{\otimes n}\widetilde{\Sigma}^{(n)}]/d$ can be used, such that
\begin{eqnarray}
    S^{(n)}_{\psi}(P_S;\mathcal{A}) = \frac{1}{1-n}\log(\tr[ X_{+1}^{P_S}\widetilde{\Sigma}_\psi^{(n)}])
\end{eqnarray}

\subsection{The conditional replica kernel}

While the above discussion holds for any experiment, we can make further progress by assuming the environment is Markovian and the apparatuses are uncorrelated as in Fig.~\ref{fig2}.
In this case, the dynamics of the unnormalized conditional replica tensor $\widetilde{\Sigma}^{(n)}(t)$ can be computed by a time local update applied to the $n$ replicas of the system $S$ by
\begin{eqnarray}
    \widetilde{\Sigma}^{(n)}(t+1)=\mathcal{K}_{t}\left( \widetilde{\Sigma}^{(n)}(t) \right).
\end{eqnarray}
Since $\mathcal{K}_t$ acts non-trivially only on the subsystem $S$, and not on the qubits in $S'$, it describes also the update for the reduced tensors $\widetilde{\Sigma}^{(n)}_{\infty}$ and $\widetilde{\Sigma}^{(n)}_{\psi}$.
For example, the unnormalized reduced conditional tensor for an initial pure state evolves as
\begin{eqnarray}\label{eq:kerneldef}
    \widetilde{\Sigma}^{(n)}_{\psi}(t+1)=\tr_{S'}\left[\mathcal{K}_{t}\left( \widetilde{\Sigma}_{\psi}^{(n)}(t)\otimes \mathds{1}^{S'} \right)\right].
\end{eqnarray}
\begin{widetext}
In the most general case of time dependent coupling between environment and apparatus, the kernel is given as
\begin{eqnarray}
    \mathcal{K}_{t}\left( \widetilde{\Sigma}^{(n)}(t) \right)=\tr_{\mathbb{E}\cup \mathcal{A}}[X^\mathcal{A}_{+1}\left(U_{a_t}U_{e_t}U_{s_t}\right)^{\otimes n}\widetilde{\Sigma}^{(n)}(t)\otimes \rho_{E_t}^{\otimes n}\otimes \rho_{A_t}^{\otimes n}\left(U_{s_t}^\dagger U_{e_t}^\dagger U_{a_t}^{\dagger}\right)^{\otimes n}]
\end{eqnarray}
or equivalently, this can be expressed explicitly with the purifications of the initial apparatus and environment states:
\begin{eqnarray*}
    \mathcal{K}_{t}\left( \widetilde{\Sigma}^{(n)}(t) \right)=\tr_{\mathbb{E}\cup \mathcal{A}}[X^\mathcal{A}_{+1}\left(U_{a_t}U_{e_t}U_{s_t}\right)^{\otimes n}\widetilde{\Sigma}^{(n)}(t)\otimes (\ket{\psi_{E_t}}\bra{\psi_{E_t}}\otimes \ket{\psi_{A_t}}\bra{\psi_{A_t}})^{\otimes n}\left(U_{s_t}^\dagger U_{e_t}^\dagger U_{a_t}^{\dagger}\right)^{\otimes n}]
\end{eqnarray*}
\end{widetext}

We now discuss a few important properties of the conditional replica kernel.
First is that it is a linear map between operators existing in $(\mathcal{H}_S\otimes H_{S'})^{\otimes n}\otimes(\mathcal{H}_S^{*}\otimes H_{S'}^{*})^{\otimes n}$.
Due to linearity, it can be treated as a matrix acting on the vectorized conditional replica tensor $\ket{\Sigma^{(n)}_t} \in (\mathcal{H}_S\otimes H_{S'})^{\otimes n}\otimes(\mathcal{H}_S^{*}\otimes H_{S'}^{*})^{\otimes n}$
\begin{eqnarray}
    \ket{\Sigma^{(n)}(t+1)}={\mathbb{K}}^{n}_{t}\ket{\Sigma^{(n)}(t)}
\end{eqnarray}
The second property is that if there is no apparatus, then the kernel, $\mathcal{K}^{(n)}$ is $n$ replicas of the completely positive trace-preserving map
\begin{eqnarray}
    \mathcal{\mathcal{K}}^{(1)}\left(\Sigma^{(1)}\left(t\right)\right)=\tr_{E}[U_{e_t}U_{s_t}\widetilde{\Sigma}^{(1)}(t)\otimes \rho_{E_t}U_{s_t}^\dagger U_{e_t}^\dagger ]
\end{eqnarray}
such that ${\mathbb{K}}_{t}^{(n)}={\mathbb{K}}_{t}^{(1)\otimes n}$.
Since these maps are CPTP, the norm of $\widetilde{\Sigma}^{(n)}$ does not change, and the kernel describes the evolution of both the normalized and unnormalized conditional replica tensor.
This is in contrast to experiments with an apparatus, where the entropy in the apparatus $S^{(n)}(\varnothing,\varnothing;\mathcal{A})$, grows over time such that the kernel only describes the evolution of the unnormalized conditional replica tensor.
This linearity property implies that statistical-mechanics-like models can be constructed for the dynamics of the Markovian quantum-enhanced experiment in the same way they can for unitary dynamics and CPTP maps~\cite{Buchold_free_fermions_PRX,theoryOftransitionsBao,Nahum_all_to_all_Q_trees}.
We perform such a construction for a random unitary brickwork below, albeit without explicitly referring to the kernel, ${\mathbb{K}}_{t}^{(n)}$.

Finally, we highlight that the conditional replica kernel inherits the symmetry properties of the conditional replica tensor.
Suppose that the conditional replica tensor is symmetric under the linear invertible map $\widetilde{\Sigma}^{(n)}(t) \rightarrow \widetilde{\Sigma}^{'(n)}(t)=F(\widetilde{\Sigma}^{(n)}(t))$ for all times $t$.
Then it follows that the kernel is also symmetric under the same linear transformation acting on its input and output spaces 
\begin{eqnarray*}
    \mathcal{K}^{(n)}_T(x)=F^{-1}\left(\mathcal{K}^{(n)}_T\left(F\left(x\right)\right)\right).
\end{eqnarray*}
This also implies that the constructed statistical mechanics model will also obey the same symmetries of the conditional replica tensor.

\subsection{Replica symmetry}~\label{sec:rep_syms}

The replica symmetry and its spontaneous symmetry breaking have been useful for describing a variety of phase transitions such as the spin-glass transitions~\cite{Castellani_2005}, the Anderson localization transition~\cite{RevModPhys.80.1355}, and the MIPT~\cite{Buchold_free_fermions_PRX,theoryOftransitionsBao,Nahum_all_to_all_Q_trees,Jian_Ludwig_criticality}.
The replica symmetry in all three cases is different and results in different universality classes for the three transitions.
In the following sections, we discuss how all three of these phenomena can be described within the framework of the Markovian experiment shown in Fig.~\ref{fig2} and compare the replica symmetry broken in each transition.
We begin with models that are described by unitary evolution for which the apparatus and environment are absent.
We then discuss the effect of environment coupling, and identify the minimum replica symmetry for CPTP maps, which include Markovian stochastic processes.

Next, we introduce the apparatus and show that the replica symmetry is generically reduced to the smallest subgroup for any of the experiments introduced above.
This is in contrast to unitary dynamics interspersed with measurements, for which a much larger replica symmetry is present.
To identify why this occurs, we investigate the $n=2$ conditional replica tensor and the requirements this larger replica symmetry imposes.
This leads naturally to the information exchange symmetry, which we discuss in the following section.
Finally, we note that the following discussion holds for any experiment described by the above framework, both Markovian and non-Markovian.

\subsubsection{Symmetry for unitary evolution}
In the simplest case of unitary evolution, with no apparatus and no environment, the conditional replica tensor is just the replicated pure state 
\begin{eqnarray}~\label{eq:untiarytensor}
\Sigma^{(n)}=\widetilde{\Sigma}^{(n)}=(\ket{\mathcal{U}_T}\bra{\mathcal{U}_T})^{\otimes n}
\end{eqnarray}
In this case, the independent permutations of the bras and kets are a symmetry $X_g\Sigma^{(n)}X_{g'}=\Sigma^{(n)}$.
In addition, the conditional replica tensor is Hermitian and therefore symmetric under $\Sigma^{(n)\dagger} = \Sigma^{(n)}$.
Combined, these symmetries form~\cite{Bao_sym_enriched,Nahum_all_to_all_Q_trees} the group $(S_n \times S_n) \rtimes Z_2$, where $\rtimes$ is the semi-direct product.
In the case that the time evolution is Gaussian (as for noninteracting particles in a disorder potential), the permutation groups enlarge and in a general context~(the unitary symmetry class `A', see Refs.~\cite{Ryu_2010,RevModPhys.80.1355} for more detail) the replica symmetry becomes a coset of either the compact or non-compact Lie groups in $2n$ dimensions depending on the choice of fermions or bosons respectively~\cite{RevModPhys.80.1355}.

These Lie groups are the replica symmetries that are broken for a subset of the Anderson transitions reviewed in Ref.~\cite{RevModPhys.80.1355}.
The Anderson transition is a transition in the eigenstates of a single particle Hamiltonian tuned by the strength of disorder.
At low disorder, and in $d\geq2$, the eigenstates are delocalized, while at larger disorder they are localized.
Ref.~\cite{RevModPhys.80.1355} describes this transition by studying the average properties of the products of two-time correlation functions.
We highlight that this method is also captured by the framework presented here:
These correlation functions can be computed from the conditional replica tensor in Eq.~\eqref{eq:untiarytensor} constructed with no apparatus or environment, and the system undergoing the unitary dynamics described by the disorder Hamiltonian.

\subsubsection{Symmetry for CPTP maps}
The doubled permutation symmetry $(S_n \times S_n)$, or more generally the enlarged replica symmetry due to independent rotations on the replicated bras and kets is a unique feature of quantum dynamics, which can distinguish the Anderson transition from the spin-glass transition.
With the addition of an environment, this extra symmetry is explicitly broken to a permutation between density matrices.
Consider an experiment with an environment but no apparatus.
The conditional replica tensor is now simply $n$ replicas of the Choi-state representing the CPTP channel between $\mathcal{H}_S'$ and $\mathcal{H}_S$:
\begin{eqnarray}
    \Sigma^{(n)}=\widetilde{\Sigma}^{(n)}=\tr_{E\cup E_{e}}[(\ket{\mathcal{U}_T}\bra{\mathcal{U}_T})^{\otimes n}]=\rho_{\mathcal{U}}^{\otimes n}.
\end{eqnarray}
In this case, the bras and kets of the conditional replica tensor cannot be permuted independently.
Instead, the symmetry is reduced to permutations of the replicated density matrix $\Sigma^{(n)}=X_g\Sigma^{(n)}X_g^{\dagger}$.
Ignoring Hermiticity, the symmetry group is now $S_n$.

This is the replica symmetry broken in the spin-glass transition~\cite{Castellani_2005} occurring in a disordered spin Hamiltonian $H(h)$ with $h$ parameterizing the random interactions between spins. 
The replica symmetry breaking in a spin-glass is seen by considering $n$ replicas, $(\rho_{\beta,h})^{\otimes n}$, of the Gibbs weight $\rho_{\beta,h}=e^{-\beta H(h)}$.
After disorder averaging over $h$, the $n$ replicas of the Gibbs weight can spontaneously break the $S_n$ replica symmetric below a critical temperature~\cite{Castellani_2005}.
This is also captured by the Markovian framework introduced above, where we assume that the system undergoes unitary dynamics by the Hamiltonian $H(h)$, and consider an environment which thermalizes the system to the Gibbs state.
As the system thermalizes, the conditional replica tensor for an arbitrary initial state $\Sigma^{(n)}_{\psi}$ will become proportional to the replicated Gibbs weight $(\rho_{\beta,h})^{\otimes n}$.
Thus, the Anderson transition and spin-glass transition are spontaneous-symmetry-breaking transitions of different replica symmetries of the conditional replica tensor.

\subsubsection{Symmetry for generic experiments}
Finally, we consider the replica symmetries possible in a system under interrogation with a measurement apparatus.
In this case, the trace over the apparatus includes the permutation unitary $X_{+1}^{(n)}$, which couples the replicas and prevents the conditional replica tensor, $\Sigma^{(n)}$, from factorizing across the replica spaces.
Still, $\Sigma^{(n)}$ will possesses at least a $Z_n\rtimes Z_2$ symmetry, where the $Z_2$ symmetry is again due to the Hermiticity.
The $Z_n$ subgroup is now due to the invariance of $\Sigma^{(n)}$ under conjugations compatible with the trace over $X_{+1}^{(n)}$.
Explicitly, it is the symmetry $X_g \Sigma^{(n)}X_{g}^{\dagger}=\Sigma^{(n)}$ where $X_g$ is constrained to centralize $X_{+1}$~(i.e. $X_g$ satisfies $\left[X_g,X_{+1}\right]=0$).
This constraint requires that the conjugation preserves the cyclic order defined by $X_{+1}$.
The only permutations that accomplish this are repeated applications of the cyclic shift $X^{(n)}_{+m}=\left(X_{+1}^{(n)}\right)^{m}$.
Since $X^{(n)}_{m}X^{(n)}_{l}=X^{(n)}_{(m+l)\%n}$, this group is $Z_n$.
Details of the proof that a generic experiment possesses at least the $D_n=Z_n\rtimes Z_2$ symmetry are shown in Appendix~\ref{apx:replicasym}

\subsubsection{Condition for the $(S_2\cross S_2)\rtimes Z_2$ symmetry in the $n=2$ conditional replica tensor}~\label{sec:n2sym}
Despite the addition of the apparatus, the dynamics of a system under projective measurements still possesses the full $(S_n\times S_n) \rtimes Z_2$ replica symmetry possessed by unitary dynamics~\cite{Bao_sym_enriched,Nahum_all_to_all_Q_trees}.
This is discussed in detail in Refs.~\cite{Bao_sym_enriched,Nahum_all_to_all_Q_trees}; and here we discuss the special case of $n=2$.
The case of $n=2$ is the simplest non-trivial case and thus offers a convenient starting place for understanding the replica symmetry breaking of the MIPT and its possible generalization to quantum-enhanced experiments.
For $n=2$, a generic experiment possesses the $D_2=Z_2\rtimes Z_2=S_2\rtimes Z_2$ symmetry introduced in the previous section.
The non-trivial symmetry operations for this reduced replica symmetry are
\begin{eqnarray*}
    \Sigma^{(2)\dagger}=\Sigma^{(2)}, \\
    X_{+1}\Sigma^{(2)}X_{+1}=\Sigma^{(2)},
\end{eqnarray*}
where $X_{+1}$ is the only nontrivial $S_2$ permutation (i.e swap of the two replicas).
We are therefore interested in the conditions under which a quantum-enhanced experiment preserves $(S_2\times S_2)\rtimes Z_2=(Z_2\times Z_2)\rtimes Z_2$.
This occurs when the remaining symmetry condition is also satisfied:
\begin{eqnarray}\label{eq:n2rsym}
    X_{+1}\Sigma^{(2)}=\Sigma^{(2)}.
\end{eqnarray}
This condition requires that the $n=2$ conditional replica tensor obtained by conditioning on the apparatus is the same as the one obtained by conditioning on the environment:
\begin{eqnarray*}
    X_{+1}^S\tilde{\Sigma}^{(2)}= X_{+1}^S\tr_{\mathcal{A} \cup \mathbb{E}}[X_{+1}^{S\cup \mathbb{E}\cup \mathcal{A}}(\ket{\mathcal{U}_T}\bra{\mathcal{U}_T})^{\otimes 2}X_{+1}^{\mathcal{A}}]= \\ \nonumber
    \tr_{\mathcal{A} \cup \mathbb{E}}[X_{+1}^{\mathbb{E}}(\ket{\mathcal{U}_T}\bra{\mathcal{U}_T})^{\otimes 2}]\equiv\tilde{\Sigma}^{(2)}_\mathbb{E}
\end{eqnarray*}
where, $\tilde{\Sigma}^{(2)}_\mathbb{E}$ is the conditional replica tensor conditioned on the environment. In the first equality we used $X_{+1}\ket{\mathcal{U}_T}^{\otimes 2}=\ket{\mathcal{U}_T}^{\otimes 2}$, and in the second we used $X_{+1}^2=1$. Introducing explicitly the dependence on the apparatus, $\tilde{\Sigma}^{(2)}\equiv \tilde{\Sigma}^{(2)}_\mathcal{A}$, we have proven
\begin{eqnarray}\label{eq:consum}
    \tilde{\Sigma}^{(2)}_\mathcal{A}=\tilde{\Sigma}^{(2)}_\mathbb{E} \leftrightarrow (S_2\cross S_2)\rtimes Z_2.
\end{eqnarray}
In the MIPT, this $(S_2\cross S_2)\rtimes Z_2$ replica symmetry is spontaneously broken~\cite{Bao_sym_enriched,Nahum_all_to_all_Q_trees}.
Therefore, to generalize the MIPT we search for a quantum enhanced experiment which also has this replica symmetry and can spontaneously break it.
The condition~\eqref{eq:consum} then states that for a such an experiment, it is necessary and sufficient that the conditional replica tensors conditioned on the environment and apparatus are equivalent.
This is the basic motivation for the information exchange symmetry relating the information in the environment to that in the apparatus.

\subsubsection{Generalization to $D_n\rtimes Z_2$}
Below we will discuss how to guarantee Eq.~\eqref{eq:n2rsym}, which is a necessary and sufficient condition for the $(S_2\times S_2)\rtimes Z_2$ MIPT replica symmetry for the $n=2$ theory.
Doing so, we will identify the local unitary exchange~(LUE) symmetry which will also guarantee an additional generator of the replica symmetry for the $n\geq2$ theories.
In Appendix~\ref{apx:replicasym}, we show that this additional generator raises the $D_n$ symmetry of a generic experiment to a $D_n\rtimes Z_2$ symmetry.
In Sec.~\ref{sec:ssb} it is demonstrated how both the MIPT and the brickwork quantum-enhanced experiment discussed below spontaneously break this $D_n\rtimes Z_2$ symmetry.
Finally, note that $D_2\rtimes Z_2=(S_2\cross S_2)\rtimes Z_2$ such that the $n=2$ limit is consistent.

\begin{figure*}[t]
  \centering
      \includegraphics[width=5.1in]{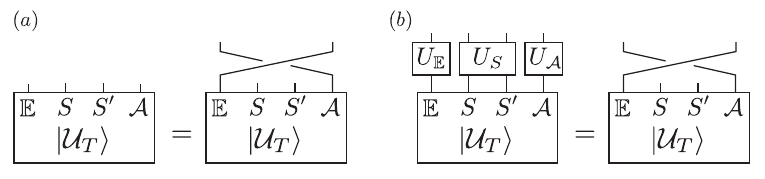}
      \caption{(a) AEE and (b) LUE symmetries for the Choi state $\ket{\mathcal{U}}$ in the case that the initial system state is not specified.}
  \label{fig:AEESNE}
\end{figure*}
\section{The Information Exchange Symmetry}~\label{SM:SNEsym}

Above, we showed that for the $n=2$ replica theory, a necessary and sufficient condition to guarantee the full $(S_2\cross S_2)\rtimes Z_2$ replica symmetry was an equivalence between the conditional replica tensor conditioned on the environment and the one conditioned on the apparatus, $\tilde{\Sigma}^{(2)}_{\mathcal{A}}=\tilde{\Sigma}^{(2)}_\mathbb{E}$.
Since this condition is regarding this abstract quantity, we now search for a unitary symmetry which guarantees that conditioning on the apparatus is equivalent to conditioning on the environment.
This is trivially guaranteed if the experiment is symmetric under exchanging the apparatus and environment
\begin{eqnarray}\label{eq:swapEA}
    \SWAP(\mathbb{E},\mathcal{A})\ket{\mathcal{U}_T}=\ket{\mathcal{U}_T},
\end{eqnarray}
which is satisfied for the projective measurements as discussed in the letter~\cite{short}.
We consider an experiment that results in this symmetry to be apparatus-environment exchange~(AEE) symmetric.

{This AEE symmetry is in fact not strictly necessary to guarantee the full $n=2$ replica symmetry. 
This is because the conditional replica tensor includes the trace over the apparatus and environment, and the permutation unitaries $X_g$ are invariant under a unitary transform $V^{\otimes n}$:
\begin{eqnarray}
    X_g=V^{\otimes n}X_g V^{\dagger \otimes n},
\end{eqnarray}
for any unitary $V$.
Therefore, we only require that the experiment is AEE symmetric up to independent unitary transforms~(see Fig.~\ref{fig:AEESNE}.) acting on the system, apparatus and environment.
}

To express this requirement we introduce the following equivalence relation on experiment states $\ket{\mathcal{U}_T}$.
\begin{definition}[LU equivalence]
    Two experiments are LU equivalent if and only if their Choi states $\ket{A}$ and $\ket{B}$ satisfy  
    \begin{eqnarray}\label{eq:swapEA'}
        \ket{A}= U_\mathcal{A}\otimes U_S\otimes U_{\mathbb{E}}\ket{B}
    \end{eqnarray}
    where $U_\mathcal{A}$,  and $U_{\mathbb{E}}$ are unitaries that act separately on the apparatus, and environment respectively, while $U_S=\prod_i U_{S_i}$ is a local unitary on the system that factorizes across the product space structure $\mathcal{H}_S=\otimes_{i\in S}\mathcal{H}_{S_i}$ of the system.
\end{definition}
Thus, the $n=2$ replica theory will posses the $D_2\rtimes Z_2$ symmetry \footnote{Note the symmetry relation~\eqref{eq:n2rsym} will need to be modified to $U_S^{\otimes n}X_{+1}\Sigma^{(2)}=\Sigma^{(2)}$. This only affects the specific nature of the transformation, but not the symmetry group. } if the experiment with the apparatus and environment exchanged, $\SWAP(\mathbb{E},\mathcal{A})\ket{\mathcal{U}_T}$, is LU equivalent to $\ket{\mathcal{U}_T}$.

This weaker requirement is expressed as a symmetry where again the symmetry transformation is the exchange of the apparatus and environment, but now only LU equivalence is required. We therefore call it LU exchange (LUE) symmetry. Since LU equivalence allows the exchange of the apparatus and environment to change the state, the LUE symmetry is not a symmetry of the state. In contrast, the symmetry transform is not allowed to transfer information between the system, apparatus, and environment. Concretely, the LU equivalence requires the eigenvalues of the reduced density matrix on the system, apparatus, or environment to be the same after the exchange of the apparatus and environment. Thus, the LUE symmetry is a symmetry of the entanglement spectrum of the reduced state of each subsystem (apparatus, environment, and system). Since the entanglement spectrum determines the R\'enyi and von Neumann entropies,  the LUE symmetry can also be considered a symmetry of the information each subsystem has about the other.

As an example, consider the noisy-transduction operation shown in Fig.~\ref{fig:PM_NT}, and imagine that the qubits $S_a$ and $S_b$ are initially in a Bell pair with qubits $R_a$ and $R_b$ respectively.
After applying the noisy transduction to this state, the apparatus and environment will have an equal amount of information about the qubits $R_a$ and $R_b$: The apparatus knows about the state on $R_b$, and the environment knows about the state on $R_a$.
Still, the state is not symmetric under the exchange of the environment and apparatus~(the AEE symmetry is not satisfied): After the exchange, the apparatus will know about $R_a$ instead of $R_b$.
Instead, the environment and apparatus are only symmetric under the exchange after a swap unitary is applied to $R_a$ and $R_b$.
Thus, while this simple experiment doesn't satisfy the AEE symmetry, it does satisfies the LUE symmetry.

When the LUE symmetry is satisfied, the $n^{th}$ replica theory satisfies a $(Z_n\rtimes Z_2)\rtimes Z_2=D_n\rtimes Z_2$ symmetry.
This is proven in Appendix~\ref{apx:replicasym}.
We also demonstrate there that this replica symmetry guarantees that the $n^{th}$ conditional replica tensor satisfies:
\begin{eqnarray}~\label{eq:sneS}
    S^{(n)}(P_S,P_{S'};\mathcal{A}) = S^{(n)}(P_S,P_{S'};\mathbb{E}),
\end{eqnarray}
and similarly, the von Neumann entropy satisfies $ S(P_S,P_{S'};\mathcal{A}) = S(P_S,P_{S'};\mathbb{E})$.
As a reminder, $P_S$ and $P_{S'}$, are subsystems of the initial and finial states of the system, respectively~(notation is defined in section~\ref{SM:replica} after Eq.~\eqref{eq:PSnSpdef}).

Below we find it useful to consider this relation as a symmetry, and to do so we introduce the following equivalence class
\begin{definition}[Informationally Equivalent]
    Two experiments are informationally equivalent if their Choi states $\ket{A}$ and $\ket{B}$ yield R\'enyi entropies that satisfy
    \begin{eqnarray}
        S_A^{(n)}(P_S,P_{S'};\mathcal{A}) = S_B^{(n)}(P_S,P_{S'};\mathcal{A}),
    \end{eqnarray}
    for all partitions $P_S\subset S$, for all partitions $P_{S'}\subset S'$ and for all R\'enyi indices $n$.
\end{definition}
Thus, the $D_n\rtimes Z_2$ replica symmetry, LUE symmetry, and AEE symmetry all guarantee the experiment is informationally equivalent to the experiment obtained by exchanging the apparatus and environment. 
We consider an experiment satisfying such a property as obeying the information exchange~(IE) symmetry.
The symmetry can also be constructed for systems initialized in a pure state, in which case the symmetry requires $S_{\psi}^{(n)}(P_S;\mathcal{A}) = S^{(n)}_{\psi}(P_S;\mathbb{E})$.

\subsection{Entanglement properties of IE symmetric experiments}\label{sec:entprops}
The IE symmetry guarantees a few important properties of the conditional entropies which relate $S(P_S|\mathcal{A})$ to a probe of entanglement~\cite{RevModPhys.81.865}. 
The first property is that the conditional entropy of a subsystem $P_S\cup P_{S'}$ of the final and initial systems $S\cup S'$, conditioned on the apparatus is equal to the conditional entropy of the complement subsystems: $S(P_S,P_{S'}|\mathcal{A})=S(P_S^c,P_{S'}^c|\mathcal{A})$.
This property also holds for the R\'eyni entropies:
\begin{eqnarray}~\label{eq:c}
    S^{(n)}(P_S,P_{S'};\mathcal{A}) = S^{(n)}(P_S^c,P_{S'}^c;\mathcal{A}),
\end{eqnarray}
and follows from the IE symmetry and the purity of the global state $\ket{\mathcal{U}}$.
This property is very natural for projective measurements, as the conditional state on the system is pure such that the entropy of a subsystem is equal to the entropy of its complement. 
Thus, in terms of the conditional entropies, the IE symmetry ensures that the conditional state on the system behaves like a pure state.

The second property holds only for the von Neumann entropy and is due to subadditivity of von Neumann entropies~\cite{nielsen2010quantum} applied to the environment, apparatus, and system:
\begin{eqnarray*}
    S(\varnothing,\varnothing;\mathcal{A})+
    S(\varnothing,\varnothing;\mathbb{E}) \leq
    S(P_S,P_{S'};\mathcal{A})+
    S(P_S,P_{S'};\mathbb{E}) 
\end{eqnarray*}
where $\varnothing$ indicates the empty set and signifies that the corresponding Hilbert space is discarded.
In combination with the IE symmetry, it yields the positivity of the quantum conditional entropy
\begin{eqnarray}
    S(P_S,P_{S'}|\mathcal{A})\geq 0.
\end{eqnarray}

Finally, by using the IE symmetry and the property Eq.~\eqref{eq:c} we can obtain
\begin{eqnarray*}
    S(P_S,P_{S'}|\mathcal{A})&=&S(P_S^c,P_{S'}^c;\mathcal{A})-S(S,S';\mathcal{A})  \\ \nonumber 
    &\equiv& -S(P_S\cup P_{S'}|\mathcal{A}\cup P_S^c\cup P_{S'}^c) 
\end{eqnarray*}
where $S(P_S\cup P_{S'}|\mathcal{A}\cup P_S^c\cup P_{S'}^c)$ is the quantum conditional entropy of $P_S\cup P_{S'}$ conditioned on $\mathcal{A}\cup P_S^c\cup P_{S'}^c$.
Physically, this relation demonstrates the presence of entanglement between regions $P_S\cup P_{S'}$ and $P_S^c\cup P_{S'}^c\cup A$.
This is deduced by the results of Refs~\cite{horodeckiPartialQuantumInformation2005,horodeckiQuantumStateMerging2007} which explain what it means to condition on a quantum state.
In particular, they show that a negative quantum conditional entropy of magnitude $\left|S(A|B)\right|$ implies the ability to distill $\left|S(A|B)\right|$ Bell pairs between the two regions.
Since a similar relation holds for $S_{\psi}(P_x|\mathcal{A})$, volume law scaling of $S_{\psi}(P_x|\mathcal{A})$ implies an extensive number of distillable Bell pairs between $P_s$ and $P_s^c\cup A$.
For projective measurements, only classical information exists on the apparatus such that the Bell pairs can only be distilled between $P_s$ and $P_s^c$.
In contrast, a more generic probe, such as noisy transduction, will allow entanglement to spread to the apparatus.
Note that while the negative quantum conditional entropy gives the number of distillable Bell pairs, it is not an entanglement monotone because it can be both positive and negative and is not strictly zero for unentangled states.

\subsection{Comparison with global symmetries}~\label{sec:symcomp}
\begin{figure}[t]
  \centering
      \includegraphics[width=1\columnwidth]{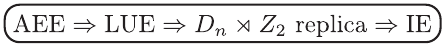}
      \caption{Relationship between the AEE, LUE, and replica symmetries, and their mutual implication for the IE symmetry expressed as a relation between von Neumann quantum conditional entropies.}
  \label{fig:implications}
\end{figure}
Fig.~\ref{fig:implications} summarizes the four symmetries which capture the idea of exchanging the information in the apparatus and environment.
The first way is the AEE symmetry, which requires the Choi state to be unchanged after exchanging the apparatus and environment.
This symmetry is generally too strong, so we introduced the LUE symmetry which is a sufficient condition on the experiment to guarantee the $D_n\rtimes Z_2$ replica symmetry.
In addition to the LU equivalence of the Choi state under the exchange of the apparatus and environment, the experiment must discard the environment degrees of freedom and condition on the apparatus degrees of freedom to guarantee the $D_n\rtimes Z_2$ symmetry.
To capture this additional requirement, we introduced the IE symmetry, which gives a necessary condition for the replica symmetry. 

For all three symmetries we introduced~(AEE, LUE and IE) the symmetry transformation is expressed as the swap unitary $\SWAP(\mathbb{E},\mathcal{A})$.
While the transformation is the same, the symmetries are distinguished by what is required to be symmetric.
In the AEE symmetry, the Choi state is required to be symmetric.
In the LUE, the structure of correlations is required to be symmetric.
For the IE symmetry only the information content, as quantified by the R\'enyi and von Neumann entropies, is required to be symmetric.

The AEE symmetry is similar to the typical use of global symmetries such as rotational symmetry, translation symmetry or the $U(1)$ charge symmetry, in which a quantum state is required to be unchanged by the symmetry operation.
In contrast, the LUE and IE symmetries only require the quantum state to be equivalent to the transformed state under some equivalence relation.
This implies that probing the symmetry requires an observable that distinguishes different equivalence classes.
Thus, a local observable on the system or apparatus cannot probe the LUE or IE symmetries.
Instead, the observable must probe correlations between the system and apparatus.

Furthermore, the symmetry transformation for the LUE or IE symmetries need not be fixed to the unitary $\SWAP(\mathbb{E},\mathcal{A})$ operation.
This is because the equivalence relation on the experiment's Choi states induces an equivalence relation on the quantum channels which act on it.
For example, a quantum channel can be considered LU equivalent to another if they can be related to each other by local unitaries applied to the apparatus, environment and system.
In this way, $\SWAP(\mathbb{E},\mathcal{A})$ need not be considered explicitly as the symmetry operation.
For the IE symmetry, one could consider any channel that transforms $S(P_S,P_{S'};\mathcal{A})\rightarrow S(P_S,P_{S'};\mathbb{E})$ as the symmetry transformation.
In this way, it is more appropriate to consider the symmetry transformation as a transformation of information as opposed to any specific transformation of a quantum state.

There is another distinction between the three symmetries introduced here and a global symmetry.
For global symmetries, the unitary transform can be viewed as either an active or a passive transformation.
As an active transformation, it describes an actual physical transformation of the system under investigation.
As a passive transformation, it describes a change of reference frame of the measurement apparatus.
For example, in the Ising model, the $Z_2$ symmetry can be viewed actively as flipping the spins, or it can be viewed passively as flipping the measurement apparatus which measures the polarization of the spins.

Neither view is appropriate for the IE symmetry.
This is because the LUE symmetry describes, in general, a transformation of both the system and the apparatus.
Furthermore, the transformation of the apparatus is not simply changing the reference frame for measuring a fixed property of the system; it changes which physical degrees of freedom are observed.
This again will prevent us from using a local observable to discuss symmetry breaking. 

\subsection{IE symmetry as an equivalence between communication tasks}~\label{sec:tasksym}
One drawback of using the von Neumann entropy or R\'enyi entropies to express a symmetry is that they are abstract quantification of correlations, and one may wonder whether if there is a more concrete expression of the symmetry.
We therefore use the fact that the von Neumann entropies can be used to quantify the capacity to perform a given set of communication task~\cite{Lloyd_1997,leditzky2016relative,Leditzky_2016}.
The IE symmetry then guarantees a symmetry between communication tasks in which the apparatus is used as a resource and one in which the environment is used as a resource. 

To make the physical nature of IE symmetry concrete, we identify one such task.
The task is to determine an arbitrary quantum state stored in a partition $P_{S'}$ of the initial system qubits by using the final state of the system, $S$, and the apparatus $\mathcal{A}$.
The coherent information
\begin{eqnarray}
    C(P_{S'}>S\cup \mathcal{A})&=&S(S,\varnothing;\mathcal{A})-S(S,P_{S'};\mathcal{A}) \\ \nonumber
    &= &S(S,\varnothing|\mathcal{A})-S(S,P_{S'}|\mathcal{A})
\end{eqnarray}
is constructed from von Neumann entropy and gives~\cite{Lloyd_1997} the maximum number of qubits encoded in $P_{S'}$ per experiment that can be decoded from $S$ and $\mathcal{A}$. 
The IE symmetry then states that the same rate can be achieved by decoding information from either the system and the environment or the system and apparatus,
\begin{eqnarray}
    C(P_{S'}>S\cup \mathcal{A})= C(P_{S'}>S\cup \mathbb{E}).
\end{eqnarray}
As we show below, the spontaneous breaking of the replica symmetry corresponds to a transition in the capacity to perform the above communication task.
We further discuss how one might consider the spontaneous breaking of the IE symmetry and also show that it corresponds to a transition in the capacity to perform the communication task.

The hierarchy of symmetries, shown in Fig.~\ref{fig:implications}, suggests interesting possibilities.
The first is that there may be an experiment for which the von Neumann entropies satisfy the IE symmetry, but for which the R\'enyi entropies don't.
Accordingly, it maybe possible to spontaneously break the IE symmetry, despite a lack of replica symmetry.
As we argue below, this transition would still show all the phenomenology of the replica symmetry breaking transition, but only for von Neumann and not R\'enyi entropies.
An even more exotic possibility is that there exists a communication task for which neither of the entropies quantifies the capacity to perform.
A symmetry between the information in the apparatus and in the environment may exist, but under a different quantifier.
This possibility highlights the need to further investigate symmetries between tasks and the possibility to spontaneously break them.

\section{A random brickwork quantum-enhanced experiment}~\label{SM:numerics}
We now present the brickwork quantum-enhanced experiment, which shows an entanglement transition.
In the accompanying letter~\cite{short}, we consider random unitaries chosen from the Clifford group and present numerical Clifford circuit simulations.
In this work we analyze the circuit in which the random unitaries are chosen from the full Haar measure.
First, we introduce the model, preparing notation needed for performing the Haar average over the random two-qudit unitaries.
Then we discuss the Haar average and highlight the different form of criticality in comparison to the MIPT.
Finally, in section~\ref{sec:ssb}, we describe the observables which detect the entanglement transition and their interpretation under replica and IE symmetry breaking.

\subsection{Noisy Transduction Circuit}
\begin{figure}[t]
    \includegraphics[width=1\columnwidth]{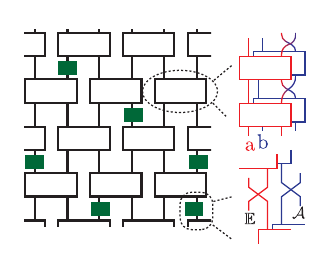}
    \caption{Example brickwork circuit composed of two-qudit unitaries and probabilistic noisy-transduction operations introduced in the letter~\cite{short}. }
    \label{fig3}
\end{figure}
Here we introduce the noisy-transduction circuit considered in the accompanying letter~\cite{short}. 
Fig.~\ref{fig3} shows the circuit again.
The system is composed of a chain of $L$ sites where at each site sits two qudits with local dimension $q$ and label $\alpha\in\{a,b\}$. 
At each step, up to $L$ environment and apparatus qudits are introduced, initialized in a pure state $\ket{0}$ and used in the noisy-transduction operation.
We label the $\alpha$ system qudit at site $x$ as $s_{x,\alpha}$, and label the environment and apparatus qudits, introduced to transduce the site $x$ at time $t$, as $e_{x,t}$ and $a_{x,t}$ respectively.
The system unitary is
\begin{eqnarray*}
    U_t&= &\prod_{\alpha \in \{a,b\},x\in 2Z_{L}}  U_{s_{x,\bar{\alpha}},s_{x+1,\alpha}}U_{s_{x,\alpha},s_{x+1,\alpha}} \text{   for $t$ even} \\ \nonumber
    U_t&= &\prod_{\alpha \in \{a,b\},x\in 2Z_{L}+1}U_{s_{x,\bar{\alpha}},s_{x+1,\alpha}}U_{s_{x,\alpha},s_{x+1,\alpha}} \text{   for $t$ odd} \\ \nonumber
\end{eqnarray*}
where the unitaries $U_{s_{x,\alpha},s_{y,\beta}}$ are two-qudit unitaries, applied to sites $s_{x,\alpha}$ and $s_{y,\beta}$, and are chosen randomly, at each time step $t$, from either the Haar measure or a distribution over the two-qubit Clifford group.
We consider periodic boundary conditions, such that on the even step, two-qudit unitaries are applied to the sites $x=L$ and $x=1$.
We also define $\bar{\alpha}=a$ when $\alpha =b$ and vice versa, such that $U_{s_{x,\bar{\alpha}},s_{x+1,\alpha}} = \SWAP(s_{x,a},s_{x,b})U_{s_{x,\alpha},s_{x+1,\alpha}}\SWAP(s_{x,a},s_{x,b})$.
Each $U_{s_{x,\alpha},s_{y,\beta}}$ forms part of a brick in the circuit shown in Fig.~\ref{fig3} a).
The full brick is shown in Fig.~\ref{fig3} b) and is formed by $\prod_{\alpha}U_{s_{x,\bar{\alpha}},s_{x+1,\alpha}}U_{s_{x,\alpha},s_{x+1,\alpha}}$, such that $U_t$ completes a single layer of bricks.

The unitary coupling the system to the environment and apparatus is given as
\begin{eqnarray}
    V_t=\prod_{x}\SWAP(s_{x,a},e_{t,x})^{\eta_{t,x}}\SWAP(s_{x,b},a_{t,x})^{\eta_{t,x}} 
\end{eqnarray}
where the variables $\eta_{t,x}\in\{0,1\}$ are independently chosen from the Bernoulli distribution with probability $p$. 
Here, $\SWAP(s_{x,a},e_{t,x})\SWAP(s_{x,b},a_{t,x})$ is the noisy-transduction operation shown in Fig.~\ref{fig3} c) applied to site $x$.

This circuit satisfies the LUE symmetry in Fig.~\ref{fig:AEESNE}.
For convenience we define $S_{\alpha}=\prod_{x}S_{x,\alpha}=\prod_x SWAP(s_{x,a},s_{x,b})$ as the swap between $a$ and $b$ qubits and $S_{AE}=\SWAP(A,E)$ as the swap between the apparatus and environment.
While the system dynamics are symmetric under the exchange of the $a$ and $b$ qudits, the noisy-transduction operation is symmetric under exchange of the $a$ and $b$ qudits and the apparatus and environment.
Therefore, the dynamics satisfy the LUE symmetry
\begin{eqnarray*}
    S_{\alpha}S_{AE}V_tU_tS_{\alpha}S_{AE}=V_tU_t.
\end{eqnarray*}

\subsection{Mapping to a model of random graphs}

In the accompanying letter~\cite{short}, we present numerical simulations of the noisy-transduction circuit for $q=2$ and the random unitaries chosen from the Clifford group.
Readers directly interested in the dynamics of information should read the letter before this section.
In this section, we consider the noisy-transduction brickwork circuit shown in Fig.~\ref{fig3} and choose the random two-qubit unitaries from the Haar distribution.
Our approach follows similar methods used in Ref~\cite{theoryOftransitionsBao}, and aims to equate the average quantum conditional entropy $\overline{S(P_S,P_{S'}|\mathcal{A})}$ to the free energy of a classical statistical mechanics model.
We first present the generic structure of this model and then derive specific weights in the following subsections.

\begin{figure}[t]
    \includegraphics[width=1\columnwidth]{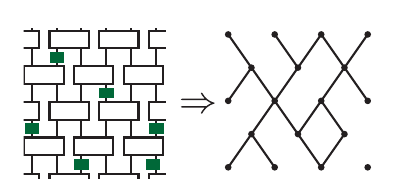}
        \caption{After performing the average over random unitaries, the replica theory for the quantum condition entropy is equivalent to the partition function for a model of random graphs.
            Each possible graph corresponds to a fixed realization of noisy-transduction locations, and the corresponding statistical weight, Eq.~\eqref{eq:partrandgraph}, is related to the average $n^{th}$ purity for the circuit with those noisy-transduction events.
        }
	\label{fig:maptographs}
\end{figure}

The approach computes the averaged quantum conditional entropy as the replica limit:
\begin{eqnarray*}
    &\overline{S(P_S,P_{S'}|\mathcal{A})}= \\ \nonumber
    &\lim_{n \rightarrow 1} \frac{1}{1-n}\left( \log\left( \overline{\tr\left[ X_{+1}^{P_S\cup P_{S'}}\widetilde{\Sigma}^{(n)} \right]} \right)-\log\left( \overline{\tr\left[\widetilde{\Sigma}^{(n)} \right]} \right) \right)
\end{eqnarray*}
where $\widetilde{\Sigma}^{(n)}$ is the unnormalized conditional replica tensor introduced above.
The statistical mechanics mapping then equates $\overline{\tr\left[ X_{+1}^{P_S\cup P_{S'}}\widetilde{\Sigma}^{(n)}\right]} $  to the partition function of a  model for random graphs~\cite{grimmett2003randomcluster}. 
The graphs are related to the tensor networks of the resulting quantum circuits for a fixed choice of noisy-transduction events $\eta_{x,t}$.
Specifically, they are the graph of the brickwork circuit, where the bricks are replaced by the vertices of the graph, and the possible edge locations correspond to the sites, $x$, and the times, $t$, at which a noisy-transduction operation might occur.
An edge is included in the graph if the noisy transduction is not performed, $\eta_{x,t}=0$, see Fig.~\ref{fig:maptographs}.
The statistical weight for a given graph is constructed from the $n^{th}$ purity
\begin{eqnarray}
    \mu^{(n)}(P_S,P_{S'},\eta_{x,t})\equiv \overline{\tr\left[ X_{+1}^{P_S\cup P_{S'}}\widetilde{\Sigma}^{(n)}_{\eta_{x,t}} \right]},
\end{eqnarray}
where the average is only performed over the two-qudit unitaries.
The partition function of the random graph model is then given by
\begin{eqnarray}~\label{eq:partrandgraph}
    Z=\sum_{\{\eta_{x,t}\}}\overline{\mu^{(n)}(P_S,P_{S'},\eta_{x,t})}\prod_{x,t}p^{\eta_{x,t}}(1-p)^{1-\eta_{x,t}}
\end{eqnarray}
which reads as the average of the $n^{th}$ purity over the choice of unitaries and noisy-transduction locations.
In this way, the weight of each graph is equal to the $n^{th}$ purity weighted by the probability of the specific realization of noisy-transduction events.
An equivalent picture holds for the brickwork circuit interspersed with measurements~\cite{theoryOftransitionsBao}.

\begin{figure}[t]
    \includegraphics[width=1\columnwidth]{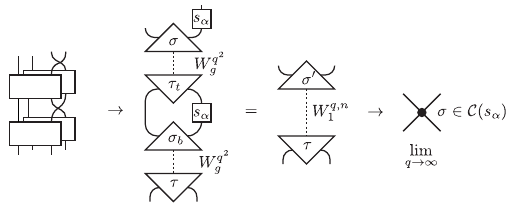}
    \caption{Steps involved in averaging over the two random two-qudit unitaries that comprise a brick. In the first step, the average is performed resulting in a sum over four permutation vectors, $\tau$, $\sigma,$ $\tau_t$, and $\sigma_b$, weighted by the Weingarten function, $W_g^{q^2}$. The $s_a$ blocks are unitaries swapping the $a$ and $b$ Hilbert spaces for each replica. In the second step, coarse-graining is performed over the $\tau_t$ and $\sigma_b$ permutations. In the final step, the large $q$ limit is taken, and the $\sigma$ and $\tau$ spins are forced to be equal and projected to the centralizer subgroup $\mathcal{C}(s_\alpha)$~(the subgroup of permutations such that $[\tau_t,s_\alpha]=0$). }
	\label{fig7}
\end{figure}
In the limit of large local Hilbert space dimension, $q\rightarrow \infty$, the random graph model becomes the random bond cluster~(RBC) model.
The RBC model is a model for random graphs with fixed vertices, and where the edges, $\gamma$, are chosen randomly from a fix set of possibilities, $\gamma \in \Gamma$.
The probability for a given graph is determined by whether an edge is included or not, $\eta(\gamma)\in \{0,1\}$, and the number of connected clusters $C$ in the resulting graph.
Each graph is then given a weight
\begin{eqnarray}
    \mu_{h,\nu}(\eta)=h^{C}\prod_{\gamma \in \Gamma}\nu^{\eta(\gamma)}(1-\nu)^{1-\eta(\gamma)}
\end{eqnarray}
parameterized by the cluster weight $h>0$, and edge probability $\nu$.
Important limits of the resulting model are $h=1$, which yields bond percolation, $h=2$ which yields the partition function of the Ising model; and an integer $h>2$, which yields the $h$-state Potts model.

In the large-$q$ limit of the MIPT, the averaged $n^{th}$ purity corresponds to the RBC model with cluster weight $h=n!$ and renormalized bond probability $\nu\neq p$ .
In this way, the MIPT is mapped to bond percolation in the $n\rightarrow 1$ and $q\rightarrow \infty$ limits. 
Here we show that, similar to the MIPT, the noisy-transduction circuit also maps to the RBC model but with $h(n)\neq n!$ being a different monotonically increasing combinatorial function of $n$.
While we are unable to perform the analytic continuation to find $\lim_{n\rightarrow 1}h(n)$, we note that $h(2)=2$ and $h(n)<n!$ for $n>2$, suggesting that the replica limit is distinct from $h\rightarrow 1$ and is thus not bond percolation.
This would be consistent with the difference in universality between the MIPT and the noisy-transduction quantum experiment observed in Clifford numerics~\cite{short}.

\subsubsection{Approach}
We now show that the random graph model~\eqref{eq:partrandgraph} for the noisy-transduction experiment yields the RBC model in the large $q$ limit, with cluster weight $h(n)<n!$.
The derivation is similar to that for measurements, but the reduction of cluster weight results from a reduction of the replica symmetry.
An overview of the approach is as follows.
We first fix the realization of noisy-transduction events $\eta_{x,t}$ and perform the Haar average.
This yields a Potts like model of $n!$ states living on a graph with the same connectivity as the graph defined by $\eta_{x,t}$. 
Taking the $q\rightarrow \infty$ limit, a subset of those states is projected out yielding a Potts model with $h(n)<n!$ living on the graph defined by $\eta_{x,t}$.
This model is the Edwards-Sokal~\cite{edwardsokalrep} representation of the RBC model whose marginal on the graph yields the $h(n)$ RBC model~\cite{grimmett2003randomcluster} with renormalized bond probability $\nu=\frac{(1-p)}{pq^{1-n}+(1-p)}$.

\subsubsection{Average of a brick}
\begin{figure}[t]
        \includegraphics[width=1\columnwidth]{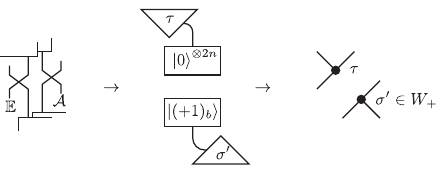}
        \caption{Noisy transduction breaks a bond in the random graph model. The center figure is the local part of the tensor network of $\overline{\tilde{\Sigma}^(n)_{\eta_{x,t}}}$ after Haar averaging. In addition to breaking the bond, noisy transduction introduces $\bra{(+1)_b}$ due to tracing out the environment and taking the expectation value of the $X_{+1}$ unitary in the apparatus.}
	\label{fig8}
\end{figure}

The average over the brickwork unitary $\overline{U_{x,x+1}^{\otimes n}\otimes U_{x,x+1}^{*\otimes n}}$ involves two steps, shown in Fig.~\ref{fig7}. First is the average over the two sets of random unitaries and second is coarse graining.
The average over the unitaries results in Eq.~\eqref{eq:wgaverage}, where the sum over permutations is not only over the replica spaces as in Ref.~\cite{theoryOftransitionsBao}, but also over the permutations involving the two sets of qudits at each site. This is because there are $2n$ copies of the unitaries being averaged, one set of $n$ from the unitaries acting on the $a$ qudits, and another from those acting on the $b$ qudits.
We therefore find
\begin{eqnarray*}
    \overline{U_{x,x+1}^{\otimes n}\otimes U_{x,x+1}^{*\otimes n}}= S_{x,\alpha}^{\otimes 2n}W_{g}^{q^{2},2n}S_{x,\alpha}^{\otimes 2n}W_{g}^{q^{2},2n}
\end{eqnarray*}
shown graphically in Fig.~\ref{fig7} where $S_{x,\alpha}$ is the swap gate applied at site $x$ between the two qudits $a$ and $b$, and $W_g^{q^2,2n}$ is the average, given in Eq.~\eqref{eq:WGdef}, over the $2n$ copies of identical unitaries and their $2n$ conjugates.
\begin{widetext}
For all $q$ and $n$ the average takes the form
\begin{eqnarray}
    \overline{U_{x,x+1}^{\otimes n}\otimes U_{x,x+1}^{*\otimes n}}=\sum_{\sigma,\tau}W^{q,n}_{\text{Brick}}(\sigma,\tau)\ket{\sigma}_x\ket{s_\alpha\sigma s_\alpha}_{x+1}\bra{\tau}_x\bra{\tau}_{x+1}
\end{eqnarray}
where $s_\alpha$ is the permutation which swaps $a$ and $b$ on all replicas, and where the weight $W_{\text{Brick}}^{(q,n)}(\sigma,\tau)$ on the two spins $\sigma$ and $\tau$ is given as
\begin{eqnarray}~\label{eq:w1qn}
    W_{\text{Brick}}^{q,n}(\sigma,\tau)=\sum_{\tau_t\sigma_b}W^{q^{2},2n}_g(\sigma,\tau_t)W^{q^{2},2n}_g(\sigma_b,\tau)Q^{q,2n}(\tau_t,\sigma_b)Q^{q,2n}(s_\alpha\tau_ts_\alpha,\sigma_b).
\end{eqnarray}
In the $q\rightarrow \infty$ limit, the dominant term in the sum is when $\tau_t=\sigma_b=s_\alpha \tau_t s_\alpha$ such that $Q^{q,2n}(\sigma_t,\sigma_t)=q^{\text{\#cycles}(\sigma_t\sigma_t^{-1})}=q^{2n}$.
The first condition $\tau_t=\sigma_b$ is similar to the condition in Refs.~\cite{theoryOftransitionsBao,Nahum_all_to_all_Q_trees} that also suppresses spin fluctuations, while the second condition $\sigma_b=s_\alpha\tau_t s_\alpha$, forces the permutation $\tau_t=\sigma_b$ to live in the centralizer $\mathcal{C}(s_\alpha)$ of $s_\alpha$~(the subgroup of permutations such that $[\tau_t,s_\alpha]=0$).
We therefore find
\begin{eqnarray}
    \lim_{q\rightarrow \infty}W_{\text{Brick}}^{q,n}(\sigma,\tau)\approx q^{-4n}\delta(\sigma=\tau \in \mathcal{C}(s_\alpha)).
\end{eqnarray}

\subsubsection{Effect of noisy transduction}
The effect of transduction is shown in Fig.~\ref{fig8}, where transduction breaks a bond in the statistical-mechanics model and injects the boundary condition from $X_{+1}^{\mathcal{A}}$.
We therefore find that the weight due to noisy transduction at the site $x$ at time $t$ is
\begin{eqnarray}
    \bra{\sigma_{x,t}}M_x(p)\ket{\tau_{x+1,t+1}}=(1-p)Q^{q,2n}(\sigma_{x,t},\tau_{x+1,t+1})\delta_{\eta_{x,t},0} 
        +p Q^{q,2n}(\sigma_{x,t},(+1)_b)\delta_{\eta_{x,t},1}
\end{eqnarray}
where $(+1)_b$ is the cyclic shift applied only to the $b$ qudits.
In the $q\rightarrow \infty$ limit, the first term becomes proportional to a $\delta$ function.
The second term acts as a field on the spin $\sigma_{x,t}$ with a maximum weight, $Q=q^{2n}$, occurring when the spin is equal to the inverse cyclic shift of the replicas on the $b$ qudits, $\sigma=(-1)_{b}$.
This element of the permutation group doesn't belong to the centralizer of the swap between the $a$ and $b$ qudits, $(-1)_{b}\notin \mathcal{C}(s_\alpha)$, and will therefore be suppressed by $W_{1}^{q,n}$~(Eq.~\eqref{eq:w1qn}).
In contrast, the identity, $\sigma=\ident$, and cyclic shift of the replicas on both $a$ and $b$ qudits, $\sigma=(-1)$, belong to the centralizer $\mathcal{C}(s_\alpha)$, but contribute a weight $Q=q^{n+1}<q^{2n}$ due to the noisy-transduction event.
\end{widetext}
In the appendix~\ref{apx:haarweight}, we prove that this is the maximum weight of $Q^{q,2n}(\sigma,(+1)_b)$ for any element $\sigma\in\mathcal{C}(s_\alpha)$.
Thus, the transduction sites create a bias toward a subset of the spin states $W_{+}\subset \mathcal{C}(s_{\alpha})$ that obtain this maximum weight~(i.e. $\sigma\in W_{+}$ if $Q^{q,2n}(\sigma,(+1)_b)=q^{n+1}$ and $\sigma \in \mathcal{C}(s_{\alpha})$).
They also create a bias toward spins outside $\sigma\in\mathcal{C}(s_\alpha)$, and at a finite $q$ this biases domains of spins outside the centralizer $\mathcal{C}(s_\alpha)$.
While this domain minimizes free energy at the boundary of a connected cluster~(where transduction sites occur), it has a free energy cost scaling with its volume.
This cost diverges with $q$ and thus, in the large-$q$ limit, these domains are projected out such that the spins are constrained to the set $W_{+}\subset \mathcal{C}(s_{\alpha})$.
These configurations yield a weight at the bonds as
\begin{eqnarray}
    &\bra{\sigma_{x,t}}M_x(p)\ket{\tau_{x+1,t+1}}= \\ \nonumber
    &(1-p)q^{2n}\delta(\sigma_{x,t},\tau_{x+1,t+1})\delta_{\eta_{x,t},0}+p q^{(n+1)}\delta_{\eta_{x,t},1}
\end{eqnarray}

\subsubsection{The RBC model}

Combing the contributions from the previous two sections yields, in the large $q$ limit, a model for multi state spins $\sigma_{x,t}=\tau_{x,t}$ on a random square lattice specified by $\eta_{x,t}$.
The number of allowed states per spin is $\left|W_{+}\right|$ and the statistical weight for a fixed configuration of bonds is given by
\begin{eqnarray}~\label{eq:esrep}
    \prod_{x,t}(1-\nu)\delta_{\eta_{x,t},1}+\nu\delta(\sigma_{x,t},\tau_{x+1,t+1})\delta_{\eta_{x,t},0}
\end{eqnarray}
where the renormalized bond probability is given as
\begin{eqnarray}
    \nu=\frac{(1-p)}{pq^{1-n}+(1-p)}\approx 1-pq^{1-n}/(1-p).
\end{eqnarray}
This is the Edwards-Sokal representation~\cite{edwardsokalrep} of the RBC model with bond probability $\nu$ and cluster weight $h=\left|W_{+}\right|$.
For integer $h(n)$, the Edwards-Sokal representation provides a mapping from the RBC model to the Potts model of $h(n)$ Potts spins, $\sigma_{x,t}$.
Taking the marginal distribution on the Potts spins yields the Potts model, while taking the marginal on the graph edges $\eta_{x,t}$ yields the RBC model.

Notice that the bond probability $\nu$ drifts towards $\nu=1$ as $q$ increases.
This also occurs in the MIPT and is because as the local Hilbert space diverges, the two-qudit unitaries tend to perfect scramblers~\cite{Claeys_2020,Pastawski_2015} which effectively protect quantum information in the $q^2\rightarrow \infty$ diverging Hilbert space of the two qudits.
Thus, it becomes harder and harder for noisy transduction, or measurements, to disrupt the spreading of quantum information.
To obtain a transition in the $q\rightarrow \infty$, we allow the transduction probability $p$ to vary with $q$ such that $\nu$ remains constant, as was done for the MIPT.

Finally, we note that the scaling is different if the initial state of the apparatus and environment is also a source of information and noise.
Specifically, we now consider the environment qudits to be initialized in a maximally mixed state and the apparatus qudits to be initialized in a maximally entangled state with an ancilla.
In this case, the weight at the bonds picks up an extra factor of $q^{1-n}$ from $Q^{q,2n}(\tau_{x+1,t+1},(+1)_b)q^{-2n}$, changing the large $q$ scaling of the renormalized bond weights
\begin{eqnarray}
    \nu'=\frac{(1-p)}{pq^{2-2n}+(1-p)}\approx 1-pq^{2-2n}/(1-p).
\end{eqnarray}
This difference may be important for investigating a $1/q$ expansion and for identifying critical exponents of the large-$q$ critical scaling exponents in the replica limit $n\rightarrow 1$.

\subsubsection{Boundary conditions}~\label{sec:BC}
The above mapping holds for infinite time and system sizes, but for finite time circuits, the initial- and final-time boundaries affect the weights of the model at the boundary.
While the boundary doesn't affect the bulk transition, the boundary conditions are determined by the partitions $P_S$ and $P_{S'}$ and are critical for determining the information quantifiers.
We discuss the final-time boundary as similar arguments hold for the initial-time boundary.
The boundary weights on the spin $\sigma_{x,T}$ are different depending on whether the qudits $x_1$ and $x_2$, on which the $(x,T)$ unitary brick acts, are in the set $P_S$ or not.
The weights are independent for each qudit $x_1$ and $x_2$ such that the weight is given as
\begin{eqnarray}
    \text{if } x_i \in P_S    &:& \quad  \braket{\ident}{\sigma_{x,T}}=Q^{q,2n}(\sigma_{x,T})  \\
    \text{else } x_i \notin P_S &:& \quad  \braket{+1}{\sigma_{x,T}}=Q^{q,2n}((+1)\sigma_{x,T}) 
\end{eqnarray}

Thus, the boundaries introduce weights which bias the Potts spins towards $(-1)\in W_+$ if $x_i\in P_S$ or $\ident\in W_+$ if $x_i\in P_S$.
The boundary weights at the initial-time boundary are the same except when the initial state is a pure state; in this case the boundary spin is free and is not biased toward any spin.

In the most general case, we are interested in the average conditional replica tensor and its symmetries.
Since the average over the Haar measure introduces a sum over the permutation vectors $\ket{\sigma}$, the average conditional replica tensor will also be a sum over a product of permutation vectors.
Explicitly, we assume $T$ is odd and write the average conditional replica tensor as
\begin{eqnarray}\label{eq:conditionalreplicatenssorpotts}
    \tilde{\Sigma} = \sum_{\{\sigma_T\},\{\sigma_0\}} Z_{\{\sigma_T\},\{\sigma_0\}} X_{\{\sigma_T\},\{\sigma_0\} }
\end{eqnarray}
where we have dropped the R\'enyi index $(n)$ and the sum is over the permutation unitaries
\begin{eqnarray*}
    X_{\{\sigma_T\},\{\sigma_0\} } = \prod_{x,x'\in 2Z_L+1}X_{\sigma_{x'-1,0},\sigma_{x',0}}X_{\sigma_{x-1,T},\sigma_{x,T}}
\end{eqnarray*}
weighted by the partition function $Z(\{\sigma_T\},\{\sigma_0\})$ for the $\left|W_{+}\right|$-spin Potts model with boundary spins fixed at $\{\sigma_T\}$ for the final-time boundary and $\{\sigma_0\}$ for the initial-time boundary.

\section{Spontaneous Symmetry Breaking}~\label{sec:ssb}
Above we showed that the R\'enyi entropies map to the free energy of the RBC model~\eqref{eq:partrandgraph}, which undergoes a phase transition tuned by the rate of noisy-transduction events.
In the RBC model, the two phases are distinguished by whether the graph is connected or not.
This view of the transition does not directly reveal the spontaneous replica symmetry breaking.
Instead, we consider the Edwards-Sokal representation~\cite{edwardsokalrep} of the free energy shown in Eq.~\eqref{eq:esrep}.
This representation works for integer cluster weight $h$ and involves $h=\left|W_+\right|$ Potts spins $\sigma_{x,t}\in W_+$, which corresponds to the subset $W_+$ of permutation group elements which acts on the replicas.
In the symmetric phase, the Potts spins $\sigma_{x,t}$ fluctuate between the different permutation group elements in $W_+$, while in the symmetry-broken phase, they condense into a specific permutation group element, determined by the boundary conditions discussed in the previous section~\ref{sec:BC}.
Thus, in this representation, replica symmetry breaking is apparent and in the next section we discuss it and its implications for the dynamics of information.
We then discuss how notions of symmetry breaking apply to the information exchange symmetry and can be used to predict the properties of the entanglement transition. 

The presentation in this section tends towards abstract results general to any model possessing the $D_n\rtimes Z_2$ replica symmetry and the accompanying Potts symmetry.
Readers interested in a concrete example can find numerical simulations of the 1D noisy-transduction circuit in the accompanying letter~\cite{short}.

\subsection{Spontaneous Symmetry Breaking of the $D_n \rtimes Z_2$ Replica Symmetry}~\label{sec:repbreaking}

In the Edwards-Sokal representation~\eqref{eq:esrep}, the transition occurs as a spontaneous breaking of the Potts symmetry.
This symmetry is directly related to the replica symmetry since, as shown by Eq.~\eqref{eq:conditionalreplicatenssorpotts}, the Potts spins correspond to the permutation unitaries that make up the conditional replica tensor.
Applying the replica permutations of the $D_n\rtimes Z_2$ replica symmetry to the conditional replica tensor, one can confirm that the Potts spins are transformed in a way that keeps the set of $W_+$ spins invariant.

In the usual symmetry-breaking framework, an experimenter would probe the expectation value of the Potts spin $\left<\sigma_{x,t}\right>$.
For replica symmetry breaking, this is not possible as it would correspond to measuring the expectation values of the permutation unitaries $X_{\sigma_{x,t}}$, which are not physical observables.
Instead, replica symmetry breaking is probed by the R\'enyi entropies which correspond to the free energy of the system with boundary conditions determined by $P_S$ and $P_{S'}$.
Similar to the case of the MIPT, these boundary conditions break the replica and Potts symmetries at the boundaries and bias the system toward a specific domain~\cite{theoryOftransitionsBao,Buchold_free_fermions_PRX,Buchhold_free_fermions_PRL}.

As discussed above, inclusion of a site in $P_S$ biases the spin to $(+1)$, while exclusion from $P_S$ biases the spin towards $\ident$. 
So long as $P_S\cup P_{S'}\neq S \cup S'$ and $P_S\cup P_{S'}\neq \varnothing \cup \varnothing$, the boundaries will bias the system towards two types of domains.
In the symmetric phase, the symmetry will be restored after a fixed length away from the boundary.
The effect of biasing multiple domains will be lost after this length such that the conditional R\'enyi entropies, $S^{(n)}(P_S,P_{S'}|\mathcal{A})$ will scale as a constant with system size.
Instead, in the symmetry-broken phase, extensive regions will be biased to a fixed domain and large domain walls will form, resulting in the free-energy~(R\'enyi entropy) scaling with the size of the domain wall.
Under appropriate choices of $P_S$ and $P_{S'}$, the domain wall length will scale with system size, resulting in volume law scaling of the conditional R\'enyi entropies.
Thus, the symmetric phase corresponds to an area law scaling of the conditional R\'enyi entropy, while the symmetry-broken phase corresponds to volume law scaling.

Crucial to the phase transition leading to area-law v.s. volume-law scaling is that domains of $(+1)$ and $\ident$ spins could form in the symmetry-broken phase.
This can only occur if $(+1)$ and $\ident$ are equivalent under the Potts symmetry.
In the case of $n=2$, this is only possible if the conditional replica tensor is symmetric under $X_{+1}\Sigma^{(2)}=\Sigma^{(2)}$~(and also $\Sigma^{(2)}X_{+1}=\Sigma^{(2)}$).

For the MIPT, this is guaranteed because the full, $(S_n\times S_n)\rtimes Z_2$ replica symmetry is satisfied.
For a generic experiment, this condition is guaranteed by the LUE symmetry and necessitates the IE symmetry~(see the discussion in Sec.~\ref{SM:SNEsym}).
Thus, the LUE symmetry guarantees the replica symmetry subgroup that, when broken in the MIPT, leads to the transition from area-law to volume-law entanglement.
When this subgroup is broken for a generic experiment, it also leads to area v.s. volume law scaling of the R\'enyi entropies.
Taking the replica limit, one finds a similar transition in the conditional von Neumann entropies, which implies the scaling transition is an entanglement transition~(see discussion in Sec~\ref{sec:entprops}).
The main difference between projective measurements and a generic probe, is that the entanglement in the latter can spread also onto the apparatus.
In this way, replica symmetry breaking of the $D_n\rtimes Z_2$ subgroup generalizes the entanglement transition due to the breaking of the full $(S_n\times S_n)\rtimes Z_2$.

\subsection{Spontaneous Information Exchange Symmetry Breaking}~\label{sec:IEsb}
Since the replica symmetry breaking is directly connected to the IE symmetry, we investigate if the phase transition can be understood as a spontaneous symmetry breaking of the IE symmetry.
Detecting the spontaneous breaking of the IE symmetry faces a similar problem as detecting replica symmetry breaking: There is no local order parameter that is physically observable~\cite{Buchold_free_fermions_PRX,theoryOftransitionsBao,Nahum_all_to_all_Q_trees,Jian_Ludwig_criticality}.
For symmetry breaking of a global unitary symmetry, a local order parameter can be constructed by identifying an observable that is not invariant under the symmetry transformation.
The issue for both the replica, and IE symmetry relations discussed above~(see Sec.~\ref{SM:SNEsym}), is that they are not expressed as a unitary transform acting on a space accessible to either of the two experiments related by the symmetry.  
For the replica symmetry, the transformation is between multiple replicas of the experiment which are not accessible for a single run of the experiment.
While for the IE symmetry, the transformation is between the apparatus and environment and the experiment does not have access to its environment.

While this may seem exotic, it is necessary if the transition is to be basis independent, as one might expect for an information or entanglement transition.
Such a transition could not be detected by a local order parameter that transforms nontrivially under some symmetry.
Otherwise, that local order parameter could be used to define a local basis and the transition would be basis dependent.

\subsubsection{Probing spontaneous symmetry breaking without a local order parameter}

As discussed in Sec.~\ref{SM:SNEsym}, the LUE and IE symmetries are not exact symmetries of a state but are symmetries of a state up to an equivalence class.
Thus, detecting their symmetry breaking, one needs to identify quantifiers that can robustly distinguish different equivalence classes.
The R\'enyi and von Neumann entropies act as such quantifiers but are not linear observables on a single copy of the experiment.
Thus, to probe symmetry breaking with them, we introduce a procedure that generalizes the way the free energy detects replica symmetry breaking as discussed in Sec.~\ref{sec:repbreaking}.

We begin by highlighting the fact that, in the discussion above, the free energy was always symmetric under the replica symmetry.
The scaling of the domain wall energy changes in the symmetry-broken phase because the symmetry is explicitly broken on the boundary where the domain walls are injected. 
A similar phenomenon occurs in any form of spontaneous symmetry breaking.
In the example of a ferromagnetic which undergoes $Z_2$ symmetry breaking, the symmetry is only broken for a fixed sample.
If multiple independent samples are obtained, each sample will have positive and negative magnetization with equal probability.
The sign of magnetization for each sample is determined by a variety of random effects such as magnetic impurities or random fluctuations of the magnetic field in which the sample was cooled. 
In general, the way the symmetry is spontaneously broken is determined by how the symmetry is explicitly broken at distant points in space-time.

\begin{figure}[t]
	\includegraphics[width=\columnwidth]{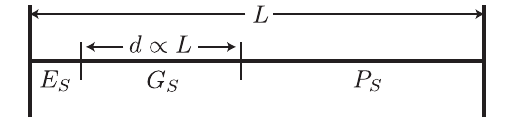}
        \caption{Example 1D geometry for probing symmetry breaking between two experiments related by a symmetry and with identical system sizes $L$.  In region $E_S$, the symmetry is explicitly broken between the two experiments, while in regions $P_S$ the symmetry is probed.  The two regions are separated by a region $G_S$ of size $d$ scaling with system size. In a symmetric phase, the symmetry will be restored in the region $P_S$ far away from the region $E_S$. While in the symmetry-broken phase, observables in the region $P_S$ will remain broken. Relevant to the entanglement transition, we have drawn the region $P_S$ with a size greater than $L/2$.}
	\label{fig:sym_break}
\end{figure}

Thus, we consider two experiments labeled by $i\in \{1,2\}$, which are symmetric to each other except in a region of space-time $E_S$.
We then probe a quantity $Q_i$ in a region of space-time $P_S$ far away from $E_S$ and consider an order parameter $O=Q_1-Q_2$.
See Fig.~\ref{fig:sym_break} for an example geometry.
The quantity $Q_i$ must transform nontrivially under the symmetry operation and distinguish between different equivalence classes.
If this is the case, then in the symmetric phase, the symmetry relation will hold asymptotically in the region $P_S$ such that $\lim_{L\rightarrow \infty}O=0$.
While in a symmetry-broken phase, the two experiments will be distinguishable far away from $E_S$ such that $\lim_{L\rightarrow \infty}O\neq 0$.

\subsubsection{The entanglement transition}\label{sec:enttrans}
We now apply this procedure to detect IE symmetry breaking.
First, we consider an experiment with the system initialized in a fixed pure state $\ket{\psi}$.
The two symmetric experiments related by the IE symmetry are one in which the experimenter has access to the apparatus and another in which they have access to the environment degrees of freedom.
To explicitly break the symmetry between the two, we consider a subset of system qudits $E_S\subset S$  at the final time boundary to be in the environment.
We can then compare the quantum conditional entropy for region $P_S$ far away from $E_S$.
The order parameter probing symmetry breaking is therefore 
\begin{eqnarray}
    O\equiv S_\psi(P_S|\mathcal{A})-S_\psi(P_S|\mathbb{E}\cup E_S)
\end{eqnarray}
Concretely, we are asking how much uncertainty about the quantum state in $P_S$ is reduced if an observer with access to the environment obtains also access to $E_S$, in comparison to an observer with access only to the apparatus $\mathcal{A}$.

For concreteness, consider a 1D system of $L$ sites and include the first $\left|E_S\right|$ qudits in $E_S$ and the last $\left|P_S\right|>L/2$ qudits in $P_S$ such that the two regions are separated by a region, $G_S$ of $L-\left|P_S\right|-\left|E_S\right|$ qudits~(see Fig.~\ref{fig:sym_break}).
The IE symmetry is satisfied asymptotically when $\lim_{L\rightarrow \infty}O=0$ and is spontaneously broken if $\lim_{L\rightarrow \infty}O$ is finite.
By using the purity of the global state of the system, environment, and apparatus we find 
\begin{eqnarray}
    O&= &S_\psi(P_S|\mathcal{A})-S_\psi(G_S|\mathcal{A})+S_\psi(E_S|\mathcal{A}) \\ \nonumber
    &= &I(P_S:E_S|\mathcal{A}),
\end{eqnarray}
where  $I(P_S:E_S|\mathcal{A})=S_\psi(P_S|\mathcal{A})+S_\psi(E_S|\mathcal{A})-S_\psi(E_S\cup P_S|\mathcal{A})$ has the form of the mutual information conditioned on the apparatus state.

If the conditional entropies obey an area law, then $\lim_{L\rightarrow \infty}O = 0$, and the system is in the symmetric phase. 
If they instead obey a volume law then $O$ depends on the relative size of regions considered.
This is because $S(A|\mathcal{A})=S(A^c|\mathcal{A})$ and volume law scaling requires that these conditional entropies scale with the size of the smaller of the two regions.
This yields, for $\left|P_S\right|>L/2$, $O\propto 2E_S$ in the volume phase.
Thus, the volume law phase corresponds to an IE symmetry-broken phase, while the area law phase corresponds to an IE symmetric phase.
Note that this argument applies to both the MIPT and the entanglement transition induced by noisy transduction.
The main difference is that for the MIPT, the entanglement is only between two regions of the system, while for a generic IE symmetric experiment, the entanglement also spreads onto the apparatus~(See discussion in Sec.~\ref{sec:entprops}).

\subsection{Spontaneous information symmetry breaking via temporal correlations}\label{sec:temptrans}

We now consider the same procedure for IE symmetry breaking, but this time we compare two experiments where the symmetry is broken at the initial-time boundary, $E_S=S'$.
We then check if the two experiments are symmetric at the final-time boundary, $P_S=S$, using a quantum condition entropy as the symmetry quantifier.
Explicitly, we take an order parameter $O=Q_1-Q_2$ with 
\begin{eqnarray}
    Q_1&=&S(S|\mathcal{A})=S(S,\varnothing;\mathcal{A})-S(\varnothing,\varnothing;\mathcal{A}) \\ \nonumber
    Q_2&=&S(S|\mathbb{E}\cup S')=S(S,S';\mathbb{E})-S(\varnothing,S';\mathbb{E}).
\end{eqnarray}
After using the property~\eqref{eq:c} and the IE symmetry we obtain
\begin{eqnarray*}
    O=C(S'>S\cup\mathcal{A})&=&S(S,\varnothing;\mathcal{A})-S(S,S';\mathcal{A}), \\ \nonumber
\end{eqnarray*}
which is the coherent information for the channel $S'$ to $S\cup \mathcal{A}$.
In the symmetric phase, the symmetry-breaking field at early times does not affect the symmetry at late times such that $\lim_{T\rightarrow \infty}C(S'>S\cup\mathcal{A})=0$.
Here we take the limit $T\rightarrow \infty$ with system size scaling with time, $L\propto T$.
In the symmetry-broken phase, the coherent information remains finite $\lim_{T\rightarrow \infty}C(S'>S\cup\mathcal{A})\neq 0$ for long times $T$.
In fact, for the MIPT, it was argued that the coherent information remains large for time exponentially long in system size~\cite{Gullans_2020}.

For a general IE symmetric experiment,  the coherent information is always positive $C(S'>S\cup \mathcal{A})\geq 0$ due to a combination of the information exchange symmetry and subadditivity~(similar to its use in Sec.~\ref{sec:entprops}).
Thus, in the symmetry-broken phase, $C(S'>S\cup \mathcal{A})>0$ and recovering quantum information initially stored in $S'$ from $S$ and the apparatus $\mathcal{A}$ becomes possible.
In contrast, in the symmetric phase, $C(S'>S\cup \mathcal{A})=0$ and communication is not possible.

This is the communication task discussed in Sec.~\ref{sec:tasksym}; it is symmetric under the exchange of the information in the apparatus and the information in the environment.
Regardless of whether the symmetry is spontaneously broken or remains intact, the same amount of information can be recovered from the late-time system state using either the apparatus or the environment.
The spontaneous symmetry-breaking transition occurs as a transition in the capacity to perform these tasks.

In the MIPT, the channel capacity~(the coherent information) between $S'$ and $S\cup\mathcal{A}$ was related to the purification dynamics of an initially maximally mixed system.
A similar relation holds in general and follows from the IE symmetry $S(S,S';\mathcal{A})= S(\varnothing,\varnothing;\mathcal{A})$ such that
\begin{eqnarray}
    C(S'>S\cup\mathcal{A})=S(S,\varnothing;\mathcal{A})-S(S,S';\mathcal{A})\\ \nonumber
    =S(S,\varnothing;\mathcal{A})-S(\varnothing,\varnothing;\mathcal{A})=S(S,\varnothing|\mathcal{A})
\end{eqnarray}
Thus, IE symmetry breaking is detected equivalently by the dynamics of the evolving quantum conditional entropy $S(S,\varnothing|\mathcal{A})$ and evolving coherent information $C(S'>S\cup\mathcal{A})$.

\section{Discussion}
In this work we have investigated requirements for observing entanglement transitions like the MIPT and found that projective measurements are not necessary.
This was accomplished by identifying other forms of interrogation, and showing that they can induce entanglement transitions generalizing the MIPT.
In the accompanying letter~\cite{short}, an explicit example is constructed in which projective measurements are replaced by the noisy-transduction operation which transfers quantum information to an apparatus and environment.
Here we identified general conditions for an interrogation operation to result in an entanglement transition.
Explicitly, we constructed a framework in which a system is coupled to an experimental apparatus that transfers information from the system to its own register of qubits.
Importantly, the framework models the environmental degrees of freedom explicitly such that the full dynamics of system, apparatus and environment can be modeled by unitary dynamics.
In this way, we showed in Sec.~\ref{SM:QEE} that both projective measurements and other forms of interrogation can be represented as unitary transformations acting on this global space.

Using this unifying framework, we identified generic conditions for an entanglement transition to occur independently of whether classical or quantum information is transferred to the qubits in the measurement apparatus.
Specifically, we identified the Information Exchange~(IE) symmetry, which requires that information transferred from the system to the apparatus is as useful as the information transferred from the system to the environment.
In Sec.~\ref{sec:entprops}, we discussed how the IE symmetry ensures entanglement structure generalizing circuits composed of projective measurements.
Next we demonstrated that the IE symmetry is spontaneously broken both in the MIPT and in the brickwork quantum-enhanced experiment introduced in the accompanying letter~\cite{short}~(See Fig.~\ref{fig3} and Sec.~\ref{SM:numerics}).
Finally in Sec.~\ref{sec:IEsb}, we showed how IE symmetry breaking generalizes the MIPT phenomenology~\cite{Chan_Nandkishore_OG,LI_Fisher_OG,Skinner_Nahum_OG,fisherreview,Gullans_2020,Vija_Self_organizedCorrection,Altman_MIPT_Codes,scalable_probes,Li_fisher_stat_codes}.
In particular, we showed how IE symmetry breaking is a transition in purification, late-time entanglement, and quantum communication capacity.

\subsection{Regarding postselection}
It is worth mentioning that these results highlight that postselection of projective measurements is not necessary for an entanglement transition similar to the MIPT.
The entanglement transition due to noisy-transduction operations does not involve projective measurements and postselection of measurement outcomes is not involved in any of the observables.
Nonetheless, there is still an unresolved obstacle towards observing the IE symmetry-breaking transition.
This obstacle of observation can now be formulated in two ways.
The first is in identifying an efficient probe of volume law entropy for regions growing with a two-dimensional space-time.
The second is to construct a decoder that can use the apparatus qubits and final-system qubits, to decode the initial quantum state on the system.

Generic solutions to the first approach require complexity similar to full state tomography of the considered region~(i.e. overhead scaling exponentially with the size of the region considered).
Similarly, we were unable to identify a simple decoder that could be used to observe the transition in communication capacity as was done in Ref.~\cite{ScrambelingTransition}.
Thus, there is still an unresolved exponential complexity towards observing the IE symmetry-breaking transition.
Nonetheless, our work emphasizes that the problem is not directly about postselection but more fundamentally about observing or making use of quantum correlation between extensively scaling regions of space-time.
Finally, we note that projective measurements are not special in this light; even if post selection is not explicitly involved, any observable of the MIPT will need to detect extensively scaling quantum correlations.

\subsection{Adaptive feedback protocols}
The framework we introduced is broadly applicable, capturing closed-system dynamics, open-system dynamics, and generic settings involving measurement apparatuses which record either classical or quantum information. 
The framework also captures experiments involving adaptive feedback considered in Refs.~\cite{Pixley_seperate_Feedback_2023_PRL,Steering-induced_2024,EntSteering,Vedika_comparing_MIPT_n_absorbing,buchhold_preselection,hauser2023continuousbreaking,MIPTofMATTER,Piroli_TrivialityQTcloseDP,Piotr_EntVAbs,sierant_EntVAbsdp1}.
Adaptive feedback protocols involve projective measurements applied at one time and unitaries applied at a future time conditioned on the outcomes of those measurements.
A simple example might involve ``preselection''~\cite{buchhold_preselection}, in which a unitary is applied locally to correct the random measurement outcome to a preselected value.
More elaborate examples~\cite{Steering-induced_2024} might determine the conditional unitary based on more complicated functions of the previous measurement outcomes.
Such feedback protocols can also show phase transitions, and with criticality observable in local order parameters without postselection.

This naturally raises the question of whether there is a connection between IE symmetry breaking and adaptive feedback transitions observable by local order parameters.
As argued previously~\cite{Pixley_seperate_Feedback_2023_PRL,Vedika_comparing_MIPT_n_absorbing}, the two types of transition are generally distinct and occur at two different critical points.
More generally, Ref.~\cite{MIPTofMATTER} argues that not only are the locations of the transition distinct, but also the natures of the transitions are distinct.
They argue that a transition observable by a local order parameter is a transition between two distinct phases of matter, while an entanglement transition such as the MIPT~\cite{Chan_Nandkishore_OG,LI_Fisher_OG,Skinner_Nahum_OG} or more generally the IE symmetry-breaking transition is not.
Their argument is that phases of matter should be efficiently observable and thus transitions in quantities that require postselection cannot distinguish phases of matter. 
Instead, in Ref.~\cite{MIPTofMATTER} they make that point that the MIPT, or more generally the IE symmetry breaking transition, should instead be considered what they call a ``classifier transition'', similar to the way computation problems can be separated by complexity classes.

Our work makes this distinction more concrete.
First we note that the adaptive feedback protocols are captured by the framework introduced in Sec.~\ref{SM:QEE}.
The way this is done depends on whether the final observables involve postselection of the measurement outcomes or not.
First consider the case that postselection is performed and in which measurement outcomes are recorded in the apparatus after the unitary feedback is applied.
In this case, the unitary implementing conditional feedback will couple the system and apparatus qubits holding measurements taken at an earlier time.
Thus, the experiment will involve a non-Markovian apparatus which can generate nonlocal temporal correlations.
Since the measurements satisfy the IE symmetry, these feedback experiments will also satisfy the IE symmetry and entanglement transitions can result.

In contrast, if measurements outcomes are discarded, our framework would treat the qubits which temporarily store the measurement outcomes as part of the environment.
In this case, there is no apparatus and the IE symmetry will generally not hold.
This is because no information will be transferred to an apparatus, but generically information will have transferred to the environment.
The only possibility for the IE symmetry is if the net result of the feedback is to prevent any information from transferring from the system to the environment.
Thus, the local observables of adaptive feedback transitions cannot observe IE symmetry breaking without this extreme condition of no information loss.

\subsection{Symmetry breaking between tasks}
Our work also highlights the nature of the MIPT as a transition between two phases distinguished by the capacity to perform a given task.
As highlighted in Ref.~\cite{MIPTofMATTER}, knowledge of the MIPT does not provide quantitative predictions for outcomes of local measurements.
Instead, it provides a quantitative assessment of the capacity to perform a given task, such as a computation or communication.
In this regard, describing the MIPT as a symmetry-breaking transition of the IE symmetry is very natural.
The IE symmetry is a symmetry between two tasks~(See Sec~\ref{sec:tasksym}): 1) decoding initial-state information with the aid of the apparatus and 2) decoding with the aid of the environment.
We then showed that the IE symmetry-breaking transition is a transition between a phase in which these tasks are impossible and one in which they are possible.

This suggests the exciting prospect of other task-based symmetries and their symmetry breaking.
The tasks discussed here are communication tasks, and one could think of investigating more general forms of symmetry breaking for symmetries relating other types of communication tasks. 
For example, there may be an information permutation symmetry breaking in which multiple agents have equivalent access to the information in the system.
One could also consider the possibility of transitions for other types of tasks such as those discussed in thermodynamics or computation.

\textbf{Acknowledgements:}
We greatly appreciate comments on the first version of the manuscript by Zack Weinstein and Gerald E. Fux. We thank Martino Stefanini, Dominic Gribben, Oksana Chelpanova, and Riccardo Javier Valencia Tortora for valuable discussions on the information exchange symmetry. We are also grateful to Zack Weinstein and Ehud Altman for stimulating ideas. We also thank Dongheng Qian for help identifying typos.
We acknowledge support by the Deutsche Forschungsgemeinschaft (DFG, German Research Foundation) – Project-ID 429529648 – TRR 306 QuCoLiMa (“Quantum Cooperativity of Light and Matter”), by the Dynamics and Topology Centre funded by the State of Rhineland Palatinate, and by the QuantERA II Programme that has received funding from the European Union’s Horizon 2020 research and innovation programme under Grant Agreement No 101017733 (’QuSiED’) and by the DFG (project number 499037529).
The authors gratefully acknowledge the computing time granted on the supercomputer MOGON 2 at Johannes Gutenberg-University Mainz (hpc.uni-mainz.de).

\appendix

\section{Review of the permutation group}~\label{sec:permgroup}
The permutation group $S_n$ is the set of $n!$ one-to-one functions acting on a finite set of $n$ elements.
An element of $S_n$ can be written by taking the $n$ elements to be the integers $1\cdots n$, and writing a permutation as the $n$-tuple $(1\rightarrow p(1), 2\rightarrow p(2), \dots n\rightarrow p(n))$ giving what each element $i$ maps to, $p(i)$.
This way of writing permutations is cumbersome, and we instead write them using cycle notation.

An $n$-cycle is any permutation $p(i)$ whose repeated application on a single element, $p(1),p(p(1)), \dots$ cycles through all elements of $i\in[1\dots n]$.
A basic property of permutations is that they can be written as a product of disjoint cycles.
Thus cycle notation writes a list of tuples describing the cycles. For example the permutation $(1\rightarrow 2, 2\rightarrow 1, 3\rightarrow 3)$ would be written as $(1 2) (3)$ or simply $(1 2)$ as the cycle $(3)$ is trivial.
The length of a cycle is the number of elements visited. 
For example, the permutation $(1 2)\in S_3$ contains a cycle of length $2$ and a cycle of length $1$, and the permutation $(1 3) (2 4) \in S_4$ contains two cycles of length $2$.
Finally, we define the composition of permutations by right multiplication: $(1 2) (2 3) = (3 1 2) = (1 3) (1 2)$ and $(p * q) (x) = (pq)(x) = p(q(x))$ where $x$ is an element of the set the permutations $p$ and $q$ act on. 

The cycle type of a permutation is a list of integers giving the lengths of the cycles in the permutation.
For example, the cycle type of $(1 2) \in S_3$ is $(2,1)$ and the cycle type of $(1 3) (2 5 4) \in S_5$ is $(3,2)$.
The cycle type identifies a permutation's conjugacy class, that is all permutations with the same cycle type are related by conjugation: There exists a $g$ such that $p'=gpg^{-1}$ if and only if $p$ and $p'$ are in the same conjugacy class.

A useful subgroup of $S_n$ is $Z_n$ generated by the cyclic shift permutation $(1 2 3 \dots n)$ which we will write as $(+1)$.
The elements can then be written as $(+m)=(+1)^m$ since composition of these elements obeys the multiplication rules of $Z_n$: $(+m)(+k)=(+(m+k)\%n)$.
We will also explicitly write the identity permutation as $\mathds{1}$ and the inverse permutations as $(-m)=(+1)^{n-m}$.

\section{Details on the replica symmetry in an experiment with an apparatus}\label{apx:replicasym}
\begin{figure}[t]
    \includegraphics[width=\columnwidth]{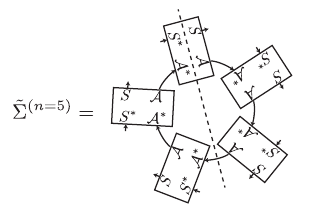}
    \caption{Graphical demonstration of the $D_5$ dihedral symmetry of the $n=5$ conditional replica tensor $\tilde{\Sigma}^{(n=5)}$. One of the lines of reflection symmetry is shown by a dashed line. Each block is the density matrix $\rho_{S,\mathcal{A}}=\Tr_{\mathcal{E}}\left[ \ket{\mathcal{U}}\bra{\mathcal{U}} \right]$, where outgoing arrows indicate the ket space and in-going arrows the bra space. Rotation symmetry follows trivially, while reflection symmetry maps out-arrows to in-arrows and therefore requires taking the Hermitian conjugate. The dihedral symmetry then follows from the Hermiticity of the density matrix $\rho_{S,\mathcal{A}}$.
    }
    \label{fig:n5sym}
\end{figure}

Above we claimed that the conditional replica tensor $\tilde{\Sigma}^{(n)}$ is guaranteed a $D_{n}$ replica symmetry regardless of the dynamics of the apparatus, system and environment.
As a reminder, the dihedral group, $D_n=Z_n\rtimes Z_2$ is the symmetry group of an $n$ sided polygon that is composed of the $Z_n$ rotation and $Z_2$ reflection subgroups.
The relation to the dihedral group can be seen in Fig.~\ref{fig:n5sym}, where we have laid out the conditional replica tensor as a contraction of the tensor $\ket{\mathcal{U}}\bra{\mathcal{U}}$.
The rotational symmetry is obvious and can be written as:
\begin{eqnarray}
    S_m(\tilde{\Sigma}^{(n)})\equiv X_{+n}^S\tilde{\Sigma}^{(n)}X_{-n}^{S}=\tilde{\Sigma}^{(n)}
\end{eqnarray}
which is proven algebraically as follows:
\begin{eqnarray}
    X_{+n}^S\tilde{\Sigma}^{(n)}X_{-n}^{S}&= &X_{+n}^S\tr\left[X_{+1}^{\mathcal{A}}(\ket{\mathcal{U}}\bra{\mathcal{U}})^{\otimes n}\right]X_{-n}^{S} \\ \nonumber
    &=&\tr\left[X_{+n}^SX_{+1}^{\mathcal{A}}X_{-n}(\ket{\mathcal{U}}\bra{\mathcal{U}})^{\otimes n}X_{+n}X_{-n}^{S}\right] \\ \nonumber
    &=&\tr\left[X_{+1}^{\mathcal{A}}X_{-n}^{\mathcal{A}\cup\mathbb{E}}(\ket{\mathcal{U}}\bra{\mathcal{U}})^{\otimes n}X_{+n}^{\mathcal{A}\cup\mathbb{E}}\right] \\ \nonumber
    &=&\tr\left[X_{+1}^{\mathcal{A}}(\ket{\mathcal{U}}\bra{\mathcal{U}})^{\otimes n}\right]=\tilde{\Sigma}^{n}
\end{eqnarray}

The reflection symmetry is less obvious and requires not only the reflection permutation, but also taking the Hermitian conjugate:
\begin{eqnarray}
    H(\tilde{\Sigma}^{(n)})\equiv X_{r}^S(\tilde{\Sigma}^{(n)})^{\dagger}X_{r}^{S}=\tilde{\Sigma}^{(n)}
\end{eqnarray}
and is proven as follows:
\begin{eqnarray}
    X_{r}^S(\tilde{\Sigma}^{(n)})^{\dagger}X_{r}^{S}&=&X_{r}^{S}\tr\left[(X_{+1}^{\mathcal{A}}(\ket{\mathcal{U}}\bra{\mathcal{U}})^{\otimes n})^{\dagger}\right]X_{r}^{S} \\ \nonumber
    &=&X_{r}^{S}\tr\left[(\ket{\mathcal{U}}\bra{\mathcal{U}})^{\otimes n}X_{-1}^{\mathcal{A}}\right]X_{r}^{S} \\ \nonumber
    &=&X_{r}^{S}\tr\left[X_{r}(\ket{\mathcal{U}}\bra{\mathcal{U}})^{\otimes n}X_{r}X_{-1}^{\mathcal{A}}\right]X_{r}^{S} \\ \nonumber
    &=&\tr\left[(\ket{\mathcal{U}}\bra{\mathcal{U}})^{\otimes n}X_{r}^{\mathcal{A}}X_{-1}^{\mathcal{A}}X_{r}^{\mathcal{A}}\right] \\ \nonumber
    &=&\tr\left[(\ket{\mathcal{U}}\bra{\mathcal{U}})^{\otimes n}X_{+1}^{\mathcal{A}}\right] \\ \nonumber
    &=&\tilde{\Sigma}^{(n)}
\end{eqnarray}
where $X_r=X_{r}^{-1}=X_{r^{-1}}$ is the reflection permutation unitary which alone generates a $Z_2$ subgroup $X_r^2=1$.
Note that neither reflection, nor Hermiticity is guaranteed alone. An example experiment that does not obey reflection or Hermiticity is simply the transduction of a single qubit, in which the initial state of the apparatus qubit $A$ is initially in a Bell pair with $A_c$.
This experiment has a conditional replica tensor $\tilde{\Sigma}^{(n)}\propto X_{+1}^n$, which under reflection $X_rX_+1X_r=X_{-1}$ is only symmetric when $n=2$ and $X_{-1}=X_{+1}$.
Similarly, the Hermitian conjugate is also not a symmetry $X_{+1}^{\dagger}=X_{-1}\neq X_{+1}$ for $n>2$.

\subsection{Replica symmetry guaranteed by the LUE symmetry}
When the LUE symmetry is satisfied the conditional replica tensor is also invariant under the permutation
\begin{eqnarray}~\label{eq:z2subgen}
    E(\tilde{\Sigma}^{(n)})\equiv X_{r}^SX_{+1}^{S}\tilde{\Sigma}^{(n)}X^S_{r^{-1}}=\tilde{\Sigma}^{(n)}
\end{eqnarray}
This is again a type of reflection, but with the addition of a cyclic shift $X_{+1}^S$ applied only to the ket space.
Similar to the above, the transformations $E$ and $S_1$ generate the dihedral group $D_n$, where $E$ performs reflections instead of $H$.
It is harder to visualize this symmetry as a geometric transformation of the conditional replica tensor, and instead we prove it algebraically:
\begin{eqnarray}~\label{eq:calc}
    & &X_{r}^SX_{+1}^{S}\tilde{\Sigma}^{(n)}X^S_{r^{-1}} \\ \nonumber
    &=&X_{r}^SX_{+1}^{S}\tr\left[X_{+1}^{\mathcal{A}}(\ket{\mathcal{U}}\bra{\mathcal{U}})^{\otimes n}\right]X^S_{r^{-1}} \\ \nonumber
    &= & X_{r}^SX_{+1}^{S}\tr\left[X_{+1}^{\mathcal{A}}X_{-1}X_{r^{-1}}(\ket{\mathcal{U}}\bra{\mathcal{U}})^{\otimes n}X_{r}\right]X^S_{r^{-1}}\\ \nonumber
    &= & \tr\left[X_{-1}^{\mathbb{E}}X_{r^{-1}}^{\mathcal{A}\cup\mathbb{E}}(\ket{\mathcal{U}}\bra{\mathcal{U}})^{\otimes n}X_{r}^{\mathcal{A}\cup\mathbb{E}}\right]\\ \nonumber
    &= & \tr\left[X_{+1}^{\mathbb{E}}(\ket{\mathcal{U}}\bra{\mathcal{U}})^{\otimes n}\right]\\ \nonumber
    &= & \tilde{\Sigma}^{(n)}
\end{eqnarray}
where in the last equality we used the LUE symmetry. Notice, that for $n=2$ the refection permutation is equivalent to the cyclic shift $X_r=X_{+1}$ such that the permutation $E(\tilde{\Sigma}^{(n)})$ is the final generator for the $S_2\cross S_2\rtimes Z_2$ symmetry as discussed in Sec.~\ref{sec:n2sym}.

Thus, when the LUE symmetry is satisfied, the conditional replica tensor guarantees three symmetric transformations: the one due to information exchange $E$, the rotation $S_1$, and the reflection with Hermitian conjugate $H$.
We now show that these transformations generate the group $D_n\rtimes Z_2$.
We will use the notation that right multiplication is the composition $ES_1H(\tilde{\Sigma})=E(S_1(H(\tilde{\Sigma})))$.
First we note that $S_1$ and $H$ generate the dihedral group as discussed above.
Furthermore, a few lines of algebra prove that $S_1=HEHE$, and the group presentation is given as $\mathcal{G}\equiv\left< H, E| H^2, E^2, (HE)^{2n}\right>$.
To prove the group is the semidirect product of $D_n$ and $Z_2$, we show that the dihedral group $D_n(S_1,H)$ generated by $S_1$ and $H$ is normal in $\mathcal{G}$.
Since this dihedral group is normal to itself, we only need to prove that $EgE=g$ for all elements $g$ in the dihedral group.
It is sufficient to do so for the generators
\begin{eqnarray}
    EHE=HS_1\in D_n(S_1,H)
\end{eqnarray}
and
\begin{eqnarray*}
    ES_1E=EHEHEE=EHEH=HS_1H\in D_n(S_1,H).
\end{eqnarray*}
Thus, $D_n(S_1,H)$ is a normal subgroup of $\mathcal{G}$.
Trivially, $E$ generates a $Z_2$ subgroup of $\mathcal{G}$ such that $\mathcal{G}=D_n\rtimes Z_2$.
Note that the argument holds using the dihedral group generated by $S_1$ and $E$ and the $Z_2$ subgroup generated by $H$.

\subsection{$D_n\rtimes Z_2$ replica symmetry guarantees IE symmetry}
Give the above $D_n\rtimes Z_2$ replica symmetry, it is easy to show that the conditional R\'enyi entropies satisfy the IE symmetry.
The replica symmetry guarantees
\begin{eqnarray}
    X_{r}^SX_{+1}^{S}\tilde{\Sigma}^{(n)}X^S_{r}=\tilde{\Sigma}^{(n)}.
\end{eqnarray}
Using all but the last equality in Eq.~\eqref{eq:calc}, we obtain that the conditional replica tensor conditioned on the apparatus is the same as the one conditioned on the environment $\tilde{\Sigma}^{(n)}=\tilde{\Sigma}^{(n)}_{\mathbb{E}}$.
From the definition of the conditional replica tensor, the IE symmetry for the R\'enyi entropies follows.
Assuming the replica limit is valid, the IE symmetry will also hold for the von Neumann entropy.

\section{Large $q$ projection to $W_{+}$ by noisy transduction}~\label{apx:haarweight}
In the Haar-average Potts-like model discussed above, noisy transduction adds a statistical weight of
\begin{eqnarray}~\label{eq:weight}
    Q^{q,2n}(\sigma_{x,t},(+1)_b))=q^{\text{\# cycles}(\sigma_{x,t}(-1)_b)}
\end{eqnarray}
In the large-$q$ limit, the average over unitaries project $\sigma_{x,t}\equiv \sigma$ to the centralizer subgroup $\mathcal{C}(s_{\alpha})$ of permutations which conjugate $s_{\alpha}$ to itself, $gs_\alpha g^{-1}=s_{\alpha}$.
Thus, the permutations that maximize Eq.~\eqref{eq:weight} in the large $q$ limit can only be chosen from that subgroup.
We now show that the maximum obtained is $q^{n+1}$.
Equivalently, we show that the maximum number of cycles of the permutation $(-1)_b\sigma$ is $n+1$ when $\sigma$ is chosen from $\mathcal{C}(s_{\alpha})$.

We first break $\sigma$ into its cycles and transpositions and discuss the constraint given by $\sigma \in \mathcal{C}(s_{\alpha})$.
Then we determine how in general a transposition changes the number of cycles in a given permutation under left multiplication.
Next we determine sequentially how each transposition in $\sigma$ changes the number of cycles in $(-1)_{b}$ to bound the number of cycles in $(-1)_b\sigma$.
In what follows it is useful to consider a set of elements $\alpha_i$, with $\alpha \in (a,b)$ and $i\in (1,n)$ which the permutations act on.

A transposition is a permutation which swaps only two elements, which as a whole generate the permutation group.
A cycle of length $|c|$ is composed of $|c|-1$ transpositions, and since different cycles act on different elements, they commute with each other.
Therefore, we can write
\begin{eqnarray}
    \sigma=\prod_j^{C_\sigma} c_j = \prod_{j=1}^{C_\sigma}\prod_{i=1}^{\left|c_j\right|}T_{j,i}
\end{eqnarray}
where $C_\sigma$ is the number of cycles in $\sigma$, the product over $i$ enumerates the cycles, and the second product enumerates the transposition $T_{j,i}$ composing each cycle $c_j$.

Since $\sigma$ is symmetric under exchange of the $a$ and $b$ replicas, each cycle must either be symmetric or have a symmetric partner contained in $\sigma$.
Similarly, a transposition in $\sigma$ is also either symmetric under exchange of the $a$ and $b$ replicas or has a symmetric partner (either in the same cycle or in the partner cycle).
The transposition of the first type can be written as $s_{\alpha,i}=(a_ib_{i})$, while the transpositions of the second type are generally $(\alpha_i\beta_j)$ with their symmetric partner equal to $(\bar{\alpha}_i\bar{\beta}_j)$. Here the overbar denotes inversion between $a$ and $b$.
Finally, we note that a symmetric cycle can be written as two independent cycles which are symmetric partners multiplied by a symmetric transposition $s_{\alpha,r_j}$,
\begin{eqnarray}
    c_j=s_{\alpha,r_j}c_{j,1}c_{j,2} =s_{\alpha,r_j}c_{j,2}c_{j,1}.
\end{eqnarray}
This is proved below.

We can therefore write the permutation with $n_s$ symmetric cycles and $(C-n_s)$ asymmetric cycles as
\begin{eqnarray}
    &\sigma=\prod_{j=1}^{n_s}s_{\alpha,r_j}\prod_{j=1}^{(C+n_s)/2}c_{j,1}c_{j,2} \\ \nonumber
    &=\prod_{j=1}^{n_s}s_{\alpha,r_j}\prod_{j=1}^{(C+n_s)/2}\prod_{i=1}^{\left|c_{j,1}\right|}T_{1,j,i}T_{2,j,i}
\end{eqnarray}
where $T_{1,j,i}=s_\alpha T_{2,j,i}s_{\alpha}$ are the $i^{th}$ transpositions of the cycles $c_{j,1}$ and $c_{j,2}$ respectively.
In the following, we will assume $j=1\dots n_s$ label the symmetric cycles.  This can always be chosen since the cycles $c_j$ are all independent and commute with each other.

\begin{figure}[t]
    \includegraphics[width=\columnwidth]{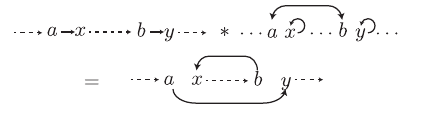}
    \caption{Effect of composing a permutation, $\sigma$ with a transposition, $T$ when the transposition acts on elements within a cycle: The cycle is broken into two cycles.
        The equality is shown by simply evaluating the composition $\sigma(T(z)$ ~(i.e $\sigma(T(b))=x$). Generally is achieved by taking the dashed lines to represent the action of the cycle on other elements in a cycle of length greater then $4$. The case for cycles of length less the $4$ can also be easily derived.}
    \label{fig:transpositioneffect}
\end{figure}

When a transposition acts on a permutation it can either increase or decrease the number of cycles by $1$.
It will increase the number of cycles if it acts within a cycle, or it will decrease the number of cycles when it acts between two cycles.
For example when the transposition acts on the identity permutation~(which has $2n$ cycles of length $1$) it combines two length 1 cycles, resulting in a permutation with $2n-1$ cycles.
Instead when the transposition acts on two elements belonging to the same cycle, two cycles will be formed and can be distinguished by which of the two element they act on.
This is proven in figure~\ref{fig:transpositioneffect}.

Now we evaluate the number of cycles in $((-1)_{b}\otimes 1_a)\sigma$ by evaluating the effect of each transposition.
That is we will keep track of the number of cycles in the permutation 
\begin{eqnarray}~\label{eq:transdef}
    &\alpha(k,m)=\\ \nonumber
    &((-1)_{b}\otimes 1_a)\prod_{j=1}^{\text{min}(k-1,n_s)}s_{\alpha,r_j}\prod_{j=1}^{k}\prod_{i=1}^{m}T_{1,j,i}T_{2,j,i}
\end{eqnarray}
where $\alpha(1,1)=((-1)_{b}\otimes 1_a)$ and $\alpha((C+n_s)/2,\left|c_C\right|)=((-1)_{b}\otimes 1_a)\sigma$.
Defining $n_{k,m}=\text{\# Cycles}(\alpha(k,m)$, we have $n_{1,1}=n+1$ and will now prove recursively that $n_{k',m'}\leq n_{k,m}\leq n+1$ for $k'>k$ and $m'>m$ when $k'=k$.

To do so, we note that the transpositions $T_{1,k,i}=(\alpha_n \beta_l)_i$ are chosen, by definition Eq.~\eqref{eq:transdef}, such that only the transpositions $T_{1,k,i}$ and $T_{1,k,i+1}$ act non-trivially on $\beta_l$.
This ensures that the action of the permutation of $\alpha(k^*,m)$ on $\beta_l$ is the same as  $((-1)_{b}\otimes 1_a)$, for $m<i$ when $k=k^*$ or when $k^*<k$.
Considering now the effect of $T_{1,k,i}$ and $T_{2,k,i}=\left(s_\alpha (\alpha_n ) s_\alpha(\beta_l)\right)$, we note that either $\beta_l$ or $s_{\alpha}(\beta_l)$ is an $a$ element invariant under $\alpha(k,m-1)$.
This guarantees that either $T_{1,k,i}$ or $T_{2,k,i}$ acts on two cycles and decrease the number of cycles, $n_{k,m}$ by one.
Since the action of the other transposition, can increase the number of cycles by at most one, the we have $n_{k,m+1}\leq n_{k,m}$.
The final step is to prove that the effect of the symmetric transposition $s_{\alpha,r_j}$ on the number of cycles is also decreasing.
This follows for a similar reason as above, since the elements $a_{(r_j)}$ and $b_{(r_j)}$ have not be acted on by a transposition, the $a$ element is guaranteed to be part of a one cycle, such that the effect is to decrease the number of transpositions.
\subsubsection{Symmetric cycle as two asymmetric cycles and a symmetric transposition}
We now prove
\begin{eqnarray}\label{eq:provethis}
    c_j=c_{j,1}s_{\alpha,r_j}c_{j,2} =s_{\alpha,r_j}c_{j,1}c_{j,2}.
\end{eqnarray}
To do so, we label the elements involved in the cycle as $x_i$ such that $c_j(x_i)=x_{i+1}$, where addition is module $\left|c_j\right|$, and we expand the permutation in transpositions as
\begin{eqnarray}
    c_j=(x_1 x_2) (x_2 x_3) \dots (x_{\left|c_j\right|-1}x_{\left|c_j\right|}).
\end{eqnarray}
First note that if $x_i=a_{r_i}$ then $x_{i+1}\neq b_{r_i}$ unless $\left|c_j\right|=2$.
Otherwise the permutation will map $a_{r_i}$ to $b_{r_i}$ but not vice-versa which is not symmetric under exchange of $a$ and $b$.
Second note that the only labeling consistent with the swap symmetry between $a$ and $b$ elements in the permutation is such $s_\alpha(x_i)=x_{(i+\left|c_j\right|/2)}$.
In this way, we can write
\begin{eqnarray}
    c_j=c_{j,1}(x_{\left|c_j\right|/2} x_{\left|c_j\right|/2+1})c_{j,2}
\end{eqnarray}
such that $s_{\alpha}c_{j,1}s_{\alpha}=c_{j,2}$.
Eq.~\eqref{eq:provethis} then follows by commuting the transposition $(x_{\left|c_j\right|/2} x_{\left|c_j\right|/2+1})$ to the left using 
\begin{eqnarray}
    c_{j,1}(x_{\left|c_j\right|/2} x_{\left|c_j\right|/2+1})c_{j,1}=s_{\alpha,r_j}=(x_1 s_{\alpha}(x_1)).
\end{eqnarray}
which follows from iterated application the group relation for transpositions $(x_1 x_2)(x_2 x_3)(x_1 x_2)=(x_1 x_3)$.

\bibliography{ref}
\end{document}